\documentclass[10pt,journal,compsoc]{IEEEtran}
\usepackage{cite}
\usepackage{ulem}
\usepackage{amsmath,amssymb,amsfonts}
\usepackage{algorithm}
\usepackage{algorithmicx}
\usepackage{graphicx}
\usepackage{textcomp}
\usepackage{xcolor}
\usepackage{subcaption}
\usepackage{booktabs}
\usepackage{multirow}
\usepackage{algpseudocode}
\usepackage{ulem} 
\usepackage[switch]{lineno} 
\usepackage[justification=centering]{caption}

\begin{document}


\title{Explainable and Transferable Adversarial Attack for ML-Based Network Intrusion Detectors}

\author{Hangsheng Zhang,
        Dongqi Han,
        Yinlong Liu,
        Zhiliang Wang,
        Jiyan Sun,
        Shangyuan Zhuang,
        Jiqiang Liu,
        Jinsong Dong}

\IEEEtitleabstractindextext{%
\begin{abstract}

Despite being widely used in network intrusion detection systems (NIDSs), machine learning (ML) has proven to be highly vulnerable to adversarial attacks.
White-box and black-box adversarial attacks of NIDS have been explored in several studies.
However, white-box attacks unrealistically assume that the attackers have full knowledge of the target NIDSs.
Meanwhile, existing black-box attacks can not achieve high attack success rate due to the weak adversarial transferability between models (e.g., neural networks and tree models).
Additionally, neither of them explains why adversarial examples exist and why they can transfer across models.

\textcolor{black}{
To address these challenges, this paper introduces ETA, an \underline{E}xplainable \underline{T}ransfer-based Black-Box Adversarial \underline{A}ttack framework. ETA aims to achieve two primary objectives: 1) create transferable adversarial examples applicable to various ML models and 2) provide insights into the existence of adversarial examples and their transferability within NIDSs.}
Specifically, we first provide a general transfer-based adversarial attack method applicable across the entire ML space.
Following that, we exploit a unique insight based on cooperative game theory and perturbation interpretations to explain adversarial examples and adversarial transferability.
On this basis, we propose an Important-Sensitive Feature Selection (ISFS) method to guide the search for adversarial examples, achieving stronger transferability and ensuring traffic-space constraints.
Finally, the experimental results on two NIDSs datasets show that our method performs significantly effectively against several classical and state-of-the-art ML classifiers, outperforming the latest baselines.
We conduct three interpretation experiments and two cases to verify our interpretation method's correctness.
Meanwhile, we uncover two major misconceptions about applying machine learning to NIDSs systems.

\end{abstract}

\begin{IEEEkeywords}
ML-based NIDSs, Adversarial attacks, Adversarial transferability, Interpretations, Cooperative game theory
\end{IEEEkeywords}}

\maketitle

\section{introduction}
\IEEEPARstart{R}{ecent} years have witnessed a steady trend of applying machine learning (ML) to detect malicious activities in network intrusion detection systems (NIDSs) \cite{mirsky2018kitsune,holland2021new,marteau2021random,sommer2010outside,chen2021adsim,fu2021realtime}.
The key enabler for this trend is the power of ML, thanks to its strong ability to detect unforeseen anomalies and high detection accuracy.
However, despite the high accuracy of state-of-the-art NIDSs, they are highly vulnerable to adversarial attacks, which evade the detection of NIDSs by applying tiny perturbations on malicious traffic.
Existing works \cite{rigaki2017adversarial,wang2018deep,hartl2020explainability,huang2020adversarial} demonstrate that even a simple adversarial attack can significantly reduce the NIDSs' accuracy.
As a result, we are concerned about using ML-based NIDSs systems for critical deployment in practice.

Therefore, we need to improve ML-based NIDSs' robustness against adversarial attacks.
The widely used approach is to construct adversarial examples (AEs) for attack evaluation.
Studies of crafting AEs against ML-based NIDSs mainly focus on white-box settings \cite{rigaki2017adversarial,wang2018deep,clements2019rallying}.
In a white-box setting, the assumption of model transparency allows complete control and access to targeted NIDSs.
Unfortunately, practical NIDSs systems do not release their internal configurations, so white-box attacks cannot be leveraged in practice.
Over the years, some researchers have developed several techniques for crafting AEs based on black-box attacks \cite{peng2019adversarial,qiu2020adversarial,alhajjar2020adversarial}.
There are two types of black-box attacks: query-based attacks and transfer-based attacks \cite{tramer2017ensemble}.
Although the query-based attacks can achieve attack success rates similar to white-box attacks, it is impractical for the attackers to obtain the detection scores in actual NIDSs.
As a result, the only way to attack NIDSs is through transfer-based attacks, which use a substitute model locally, without knowing the ML models used in NIDSs, and transfer the AEs to various ML models to evade detection.

However, applying transfer-based attacks to ML-based NIDSs faces two challenges.
First, prior transfer-based attacks within the gradient-based framework are not generic \cite{dong2018boosting,wu2020skip,wang2020unified}, focusing more on transfer across neural network architectures.
But NIDSs rely heavily on traditional ML techniques \cite{sharafaldin2018toward,patil2019designing,rashid2020machine,verma2018statistical}, such as non-differentiable tree models, and using neural networks as substitutes to attack tree models has a low success rate \cite{yang2018adversarial}.
Further, taking non-differentiable models as substitutes will not apply gradient descent directly.
Second, the underlying reasons for AEs existing and why adversarial attacks can transfer are not well understood.
It is challenging to find AEs effectively without knowing why they exist, blindly searching AEs in all features will prevent us from finding legitimate AEs, and limit the transferability of AEs.
In summary, it is not trivial to craft AEs with a high success rate under transfer-based black-box attack settings.

Our high-level goal in this paper is to develop an adversarial attack framework that is both \textbf{transferrable} and \textbf{explainable}.
Specifically, to increase the efficiency of attacks on non-differentiable models, we first train generic ensemble substitute models containing differentiable and non-differentiable base models.
As gradient-based methods cannot be applied to non-differentiable ensemble models, we employ a gradient evaluation based on zero-order optimization to craft adversarial examples (AEs). 
To further improve the transferability of the ensemble substitute model, a min-max optimization method is used.
Second, we exploit a unique insight to explain AEs and adversarial transferability. 
Based on cooperative game theory (GoGT), we calculate the importance of features and then use a perturbation-based interpretation to evaluate their sensitivity.
The results show that adversarial vulnerability directly results from important and sensitive features, which are called \textit{\textbf{non-robust features}}.
In addition, adversarial transferability is caused by two models learning the same non-robust features.
Motivated by the key insight, we propose the Important-Sensitive Feature Selection (ISFS) method to guide the search for AEs while enhancing the transferability of adversarial attacks.
To our surprise, this approach also ensures traffic-space constraints more effectively.

When examining interpretive experiments for AEs, two major misconceptions about the current use of machine learning in NIDSs systems are revealed.
First, \textbf{focusing solely on model accuracy is inappropriate}. 
Our experiments demonstrate that only using non-robust features can also train a high-accuracy model, so the perfect accuracy achieved by existing works \cite{sharafaldin2018toward,patil2019designing,rashid2020machine,verma2018statistical,zhou2020building,mirsky2018kitsune,marteau2021random} may not indicate that models found the essential characteristics of the traffic. 
Second, \textbf{choosing intrinsic features is more important than optimizing the model in current ML-based NIDSs}.
Our method is a model-agnostic (cross-model) transfer-based attack, so the non-robust features we find are more likely to be a vulnerability in feature extractors.
Any elaborate models based on non-robust features cannot learn the intrinsic traffic patterns and are easily evaded.
In summary, models that focus only on accuracy and model-centric optimization are fragile, making it hard to meet the reliability and stability required in industrial production.

Our main contributions are summarized as follows:

\begin{enumerate}
    \item We propose a generic adversarial attack architecture for ML-based NIDSs in non-interactive black-box setting. To our best knowledge, we are the first to develop a generic transfer-based adversarial attack in NIDSs.
    \item We provide a unique insight to explain the reason for the existence of AEs and adversarial transferability in ML-based NIDSs based on a combination of cooperative game theory and perturbation interpretation. Motivated by interpretation, we present a novel method Important-Sensitive Feature Selection (ISFS) to enhance transferability and satisfy traffic-space constraints.
    \item We systematically evaluate the efficiency and effectiveness of adversarial attacks against several classical and two state-of-the-art ML-based NIDSs using two NIDSs datasets and two feature sets (flow-based and packet-based). Our experimental results show that our method achieves an average attack success rate of 70\%, a 30\% improvement over four latest baselines on the datasets.
    \item We conduct three interpretation experiments (removing, using, and comparing non-robust features) and two cases to explain the causes of AEs and adversarial transferability.
    Meanwhile, we reveal two major misconceptions with the current application of machine learning to NIDSs systems in interpretive experiments.
    The findings could provide security researchers with a better understanding of model decisions and an ideal direction for developing ML-based NIDSs.
\end{enumerate}

The rest of the paper is organized as follows:
In Section \ref{sec-bg}, we provide backgrounds on ML-based NIDSs and summarize the related work.
Section \ref{sec-overview} introduces the overview of ETA design.
We describe the overall solution to craft transfer-based AEs in Section \ref{sec-eta}.
Experimental results and findings are shown in Section \ref{sec-evaluation}.
We make discussions on limitations and future works in Section \ref{sec-discuss}.
And Section \ref{sec-conclusion} concludes this study.

\section{background and related work}
\label{sec-bg}
In this section, we first introduce prior techniques for ML-based NIDSs and adversarial attacks.
Then we review some previous works most related to our paper.

\subsection{ML-based NIDSs}
Artificial intelligence technology has been utilized in NIDSs design to improve detection accuracy and efficiency. 
\subsubsection{Statistical feature-based Models}

\textcolor{black}{
Initially, the researchers focused on classical machine learning methods \cite{sharafaldin2018toward,patil2019designing,rashid2020machine,verma2018statistical} (e.g., LR, SVM, DT) applied to NIDSs, followed by an ensemble model \cite{zhou2020building}, such as Xgboost, RF and Deep Forest \cite{zhang2019deep}.
Following that, a few researchers began employing simple deep-learning models to detect malicious traffic.
For example, AlertNet \cite{vinayakumar2019deep}, DeepNet \cite{gao2020malicious}, and IdsNet\cite{zolbayar2022generating} are based on fully connected perceptions with ReLU activation functions, batch normalization to improve training performance, and dropout to prevent overfitting.}

\subsubsection{Temporal feature-based Models}
\textcolor{black}{According to the sequence of packet lengths and packet arrival intervals, flows are classified using time series models such as Markov Models and Recurrent Neural Networks.
Liu et al. \cite{liu2018mampf} proposed the Multi-attribute Markov Probability Fingerprints (MaMPF), for encrypted traffic classification, which can capture the time-series packet lengths effectively using power-law distributions and relative occurrence probabilities of all considered applications.
Liu et al. \cite{liu2019fs} applied the recurrent neural network to the encrypted traffic classification problem and propose the Flow Sequence Network (FS-Net) which is an end-to-end classification model that learns representative features from the raw flows.}
It is common to apply a sequence model to traffic classification, and this method can also be used to detect malicious traffic as well.
Fu et al. proposed Whisper \cite{fu2021realtime}, a real-time ML-based malicious traffic detection system with high throughput and accuracy based on frequency domain features.

\subsubsection{Anomaly Detection Models}
\textcolor{black}{In recent years, anomaly-based ML methods have gained popularity because they allow the detection of zero-day threats.
As a classic example, Mirsky et al.\cite{mirsky2018kitsune} developed KitNET, an online unsupervised anomaly detector based on an ensemble of auto-encoders that characterizes anomalies by using reconstruction errors.
Following that, \cite{marteau2021random} presented DiFF-RF, an ensemble approach based on random partitioning binary trees that can detect point-wise and collective anomalies.
To summarize, ML-based NIDSs play an important role in detecting malicious activities in networks.}

\subsection{Adversarial Attacks on ML-based NIDSs}
Machine learning has given rise to a new network intrusion attack, namely an adversarial attack to evade ML-based NIDSs.
Current adversarial attacks on NIDSs can be classified into three categories based on the attacker's knowledge of the target NIDSs: white-box attacks, query-based black-box attacks, and non-query-based black-box attacks.

\subsubsection{White-box Attacks}
White-box attacks were the earliest attempts at adversarial attacks against ML-based NIDSs.
Rigaki et al.\cite{rigaki2017adversarial} presented the first try to apply adversarial attacks to NIDSs from the deep learning image classification domain.
Then \cite{wang2018deep} evaluated four gradient attack algorithms in the DL-based NIDSs.
Gradient-based techniques \cite{clements2019rallying} were also used to attack Kitnet, a state-of-the-art ML-based NIDSs.

\subsubsection{Query-based Black-box Attacks}
The query-based black-box attacks require feedback from the target classifier but without any other classifier parameters.
\cite{lin2018idsgan} proposed IDS-GAN, which utilized a Generating Adversarial Network (GAN) to transform original malicious traffic into adversarial traffic, misleading NIDSs systems.
Wu et al.\cite{wu2019evading} employed deep reinforcement learning to generate adversarial traffic to deceive a target model automatically.
After this, \cite{alhajjar2020adversarial} used particle swarm optimization (PSO) and genetic algorithms (GA) to generate AEs.
From another perspective, \cite{qiu2020adversarial} proposed the model extraction technique to replicate the model and utilized saliency maps to identify the critical features for crafting AEs.

\subsubsection{Non-query-based Black-box Attacks}
Neither feedback nor knowledge of the classifier is required for non-query-based black-box attacks.
Several traffic obfuscation methods have been proposed for evading traffic detection systems.
Apruzzese et al. \cite{apruzzese2018evading} directly modified four features at random to evade botnet detectors.
To enhance the attack success rate of random mutation, Han et al.\cite{han2020practical} first applied a GAN to generate adversarial features, followed by using PSO to make traffic-space vectors close to the adversarial features.
\textcolor{black}{
Nasr et al. \cite{nasr2021defeating} attempted to identify a (blind) perturbation vector aiming to cause a DNN model to misclassify when added to an arbitrary input from the target input domain. Besides, they enforced network traffic constraints within the optimization problem by implementing remapping functions and regularizers.
Despite this, traffic-space mimicry attacks do not achieve high evasion success rates because they do not fully benefit from the vulnerabilities of ML models.
Furthermore, as the only transfer-based attack in NIDSs, \cite{alhajjar2020adversarial} that uses the deep learning model as a substitute can also not achieve a high success rate in all ML models. 
\textbf{Perhaps the substitute models cannot fully express the function and structure of the target model.
Therefore, we need to study transfer-based black-box attacks in NIDSs systematically.}}

\subsection{Transfer-based Attacks}
Transfer-based attacks have been extensively studied in other fields.
\cite{papernot2016transferability,papernotpractical} were the earliest attempts at adversarial transferability, including deep learning and machine learning.
However, they did not optimize the transferability of the models but only tested their transferability.
Transfer-based attacks are brutal because it is challenging to achieve a high success rate.
Several methods have been proposed to increase the transferability of adversarial attacks.

The momentum iterative attack (MI Attack) \cite{dong2018boosting} incorporated the momentum of gradients to boost the transferability.
The variance-reduced attack (VR Attack) \cite{wu2018understanding} used the smoothed gradients to craft perturbations with high transferability.
The skip gradient method (SGM Attack) \cite{wu2020skip} utilized the gradients of the skip connection to improve the transferability.
Since they do not examine the underlying causes of transferability, they cannot be directly applied to traffic domain.
Wang et al. \cite{wang2020unified} put forward a unified explanation for the above methods based on game theory, which found a negative correlation between transferability and game interactions.
\textbf{However, they explored only the transferability between neural network architectures and did not examine the entire machine learning space. This inspired us to apply game theory to explain and boost adversarial transferability across the whole machine learning space.
Besides, the study did not optimize the substitute model but used the base or simple ensemble models.}

\subsection{Reasons for the Existence of the Adversarial Examples and Adversarial Transferability}
\label{subsec-reason}

To explain why adversarial examples (AEs) exist and the reason for adversarial transferability, extensive studies have been devoted.
The early attempt \cite{goodfellow2014explaining} explained the AEs focused on nonlinearity and overfitting.
\cite{tabacof2016exploring} demonstrated that adversarial images appear in large areas of pixel space.
\cite{moosavi2017universal} explained the AEs in terms of universal perturbations, which arise because of significant geometric correlations among the high-dimensional decision boundary of classifiers.
On the other hand, \cite{liu2016delving} examined the transferability of targeted AEs and found that non-targeted attacks were easy to transfer, but targeted attacks were incredibly difficult.
Following that, Wu et al.\cite{wu2018understanding} and Demontis et al.\cite{demontis2019adversarial} examined factors that influence transferability, such as network architectures, model capacity, and gradient alignment.
Among all the work we compared, the most similar work was \cite{ilyas2019adversarial}, which demonstrated that AEs and adversarial transferability could be directly attributed to non-robust features.
\textbf{The theory of non-robust features is ideally suited for explaining AEs in NIDSs owing to the feature-centric training process. 
But contrary to the well-defined robust or non-robust features in image scenarios, it is not easy to determine which features are non-robust in NIDSs.}
\section{Overview}
\label{sec-overview}
\begin{figure*}[h]
    \centering
    \begin{subfigure}{0.7\textwidth}
      \includegraphics[width=1\textwidth]{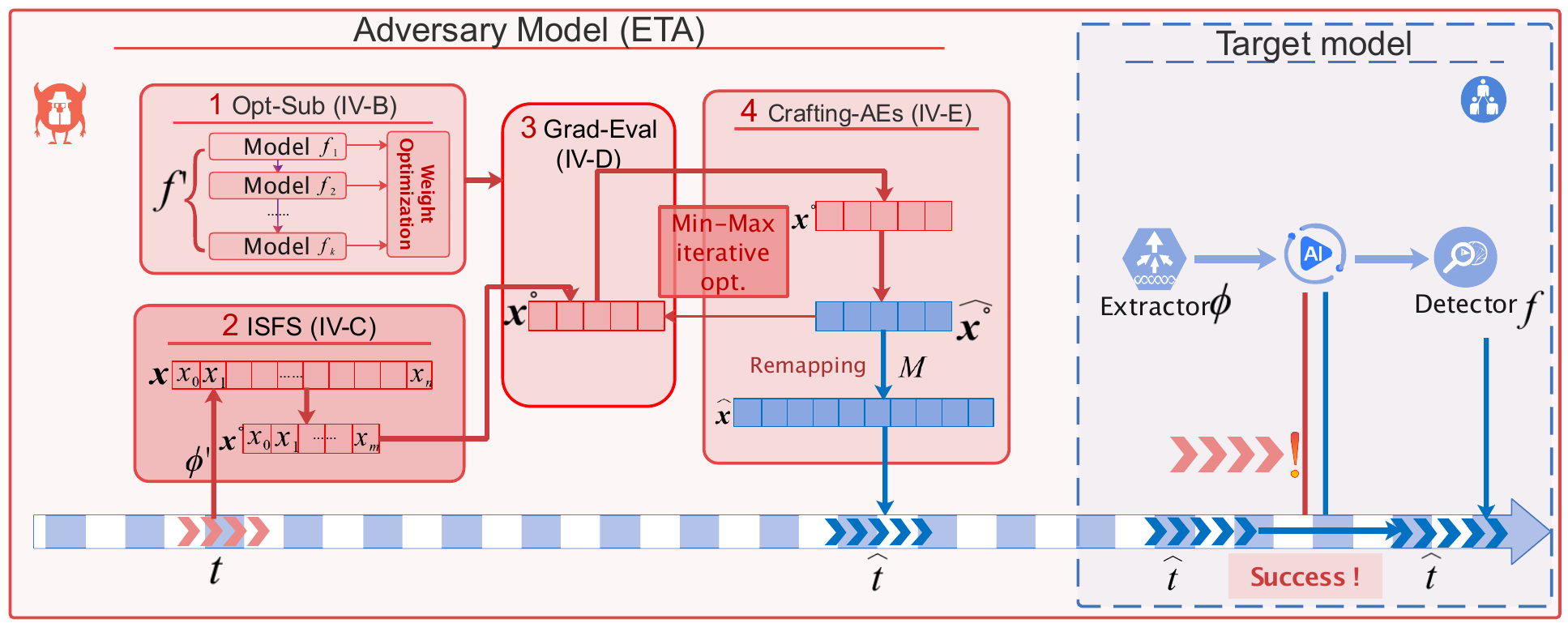}
      \caption{Adversary Model and Target Model}
    \end{subfigure}
    \begin{subfigure}{0.25\textwidth}
    \includegraphics[width=1\textwidth]{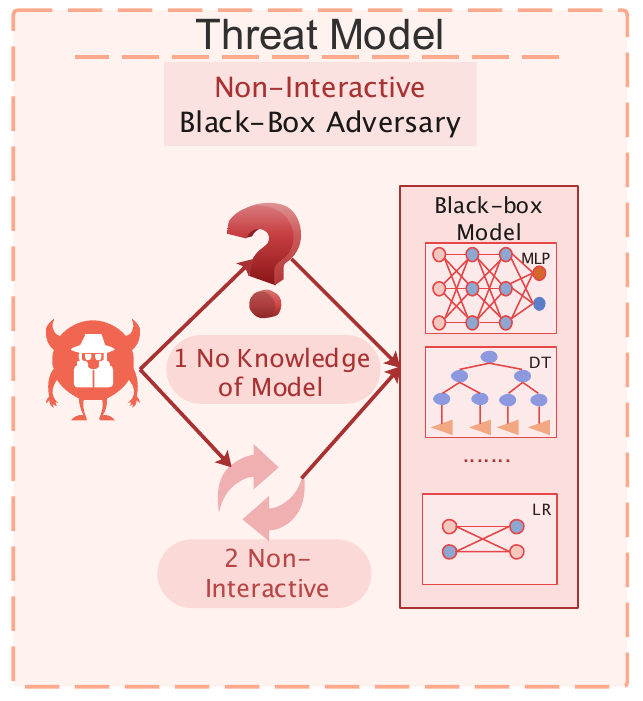}
    \caption{Threat Model}
    \end{subfigure}
    \vspace{-0.05in}
    \caption{\textcolor{black}{ Overview of system design. The left of (a) overviews ETA framework, including four core steps: 1) Optimizing the substitute model \textbf{(Opt-Sub)}, 2) Important-Sensitive Feature Selection \textbf{(ISFS)}, 3) Gradient evaluation based on zeroth-order optimization \textbf{(Grad-Eval)}, 4) Crafting AEs in domain constraints. \textbf{(Crafting-AEs)}.
    The right of (a) present the target model, which consists of a traffic capturer, feature extractor, and ML-based classifier.
    In (b) we illustrate the threat model, where the attacker does not know the target model’s architecture and cannot freely interact with the target models.}}
    \label{fig-framework}
    \vspace{-0.15in}
    \end{figure*}

\textcolor{black}{
In this section, we first present the target victim NIDS model and a more realistic adversary threat model. Then the 
consists and steps of our Explainable Transfer-based Attack (ETA) that satisfy the characteristics of the threat model are presented in general.}

\subsection{Target Model}
\textcolor{black}{
We use ML-based NIDSs as target models to examine the adversarial attack. 
To be suitable for both encrypted and unencrypted network traffic, 
the target models detect malicious traffic from network traffic patterns(e.g., time series (packet length and time interval), protocol fields), rather than payload-based features.
As shown on the right of Fig.~\ref{fig-framework}(a), the target model consists of a traffic capturer, feature extractor $\phi$, and ML-based classifier $f$. 
The captured traffic $\mathbf{t}$ is first parsed to extract feature vectors $\mathbf{x}$ by the feature extractor $\phi (\mathbf{t})$.
Then the feature vectors $\mathbf{x}$ as patterns of network traffic are fed into the ML classifier $f(\mathbf{x})$, which outputs the malicious probabilities $q$.
}



\subsection{Threat Model}
\textcolor{black}{
In the most realistic situations, the adversary has rather limited knowledge of the target model and does not freely interact with the target model. Therefore, as shown in Fig.~\ref{fig-framework}(b), the threat model is assumed as having black-box access to the target model. That is, the attacker does not know the target model’s architecture or hyper-parameters and has no feedback from the NIDS. 
In this case, the target NIDS can only be attacked by exploiting transfer-based adversarial attacks. Specifically, the attacker first constructs a substitute model locally according to his available knowledge and then against the target NIDS by countering the substitute model. Regarding the attacker's available knowledge of the target NIDS, we identify the following four elements:
}


\textcolor{black}{
\begin{itemize}
    \item \textbf{Oracle.} To avoid triggering some self-protection mechanisms of NIDS (e.g. blacklisting the attacker's IP address), it is most likely that the attacker does not interact with the target NIDS.
    \item \textbf{Detection Model.} It is less likely that attackers will be aware of the target model's architecture and parameters. As a result, this paper focuses more on cross-model transferability, i.e., transferring adversarial examples between one model (substitute model) to another model (target model).
    \item \textbf{Training Data.} It is difficult for an attacker to access the full training data used in the target NIDS model in reality. But there are many websites have exposed some malicious traffic, which is often used by security engineers to train NIDS. Thus it is reasonable to assume that the attacker is aware of partial training data of target NIDS.
    \item \textbf{Feature Extractor.} It is difficult for an attacker to fully understand what feature sets are used by the target model. But they may be able to speculate some features by their domain expertise and multiple freely available feature extractors. Therefore, it is realistic to assume that the attacker has partial knowledge of the feature sets used by the target NIDS.
\end{itemize}
}

\subsection{Overview of ETA Design}
\textcolor{black}{
Based on the above target and threat model settings, we present a general \textbf{E}xplainable \textbf{T}ransfer-based black-box adversarial \textbf{A}ttack (ETA) framework against target NIDSs. As shown on the left of Figure~\ref{fig-framework}(a), ETA includes four core parts: (1) Optimizing the substitute model, (2) Important-Sensitive Feature Selection (ISFS), (3) Gradient evaluation based on zeroth-order optimization, (4) Crafting adversarial examples (AEs) in domain constraints.}

\textcolor{black}{
Specifically, we first select a substitute dataset and use a substitute feature extractor to train a local substitute model. In particular, aiming to improve transferability, we choose and optimize an ensemble model that integrates both differentiable and non-differentiable models as the substitute model. Second, the Important-Sensitive Feature Selection (ISFS) approach based on a combination of cooperative game theory and perturbation interpretations is presented to assess the importance and sensitivity of the features and explain the reason for the existence of AEs and adversarial transferability. And then, we utilize a gradient evaluation method based on zeroth-order optimization for features selected by the ISFS method, which guides us to search for AEs towards disrupting critical and sensitive features, achieving stronger transferability.
Finally, based on the direction of the gradient evaluation, the AEs are generated iteratively and projected into a suitable region restricted by the traffic-space constraints.}

\textcolor{black}{
Each part of the ETA framework will be described in the following section in detail. Some important notations covered in this paper are listed in Table~\ref{table-notations}.}
\begin{table}[h]
    \caption{NOTATIONS IN PROBLEM FORMULATION} 
    \label{table-notations}
    \begin{tabular}{@{}ll@{}}
    \toprule
    Notations & Description                                                    \\ \midrule
    $\mathbf{t},\hat{\mathbf{t}}$         & original and adversarial traffic                               \\
    $\mathcal{T}$        & traffic-space                                                  \\
    $\mathbf{x},\hat{\mathbf{x}}$         & feature vector extracted from original and adversarial traffic \\
    $\mathbf{x}^\circ$           & Selection of feature vector\\
    $\mathcal{X}$         & feature-space                                                  \\
    $\phi,\phi'$         & targeted and substitute feature extractor                      \\
    $f,f'$         & targeted and substitute ML classifier                          \\
    $n$         & dimensionality of the original feature vector                        \\
    $m$         & dimensionality of the selection of feature vector                         \\
    $k$         & number of models in ensemble model                       \\
    $l$         & distance function between two feature vectors                  \\
    $\delta$        & perturbation added to the feature vector                       \\
    $\varTheta$         & feature-space constraints                                      \\
    $\varGamma$         & traffic-space constraints                                      \\
    $\mathcal{M}$        & remapping function                                            \\ \bottomrule
    \end{tabular}
    \end{table}

\section{System Design}
\label{sec-eta}
In this section, we begin by presenting the critical formulation for crafting transferable AEs against ML-based NIDSs. And then, we describe the overall solution to craft AEs in detail.

\subsection{The General Formulation}
\subsubsection{Definition}

\newtheorem{theorem}{Definition}{}
\begin{theorem}
\textit{\textbf{Feature Extraction.}}
A feature extraction is a function $\phi: \mathcal{T} \longrightarrow \mathcal{X} \subseteq \mathbb{R}^{n}$ that, given a traffic object $\mathbf{t}\in \mathcal{T}$, generates an $n$-dimensional feature vector $\mathbf{x} \in \mathcal{X}$, such that $\phi(\mathbf{t})=\boldsymbol{\mathbf{x}}$.
\end{theorem}

\begin{theorem}
\textit{\textbf{Traffic-space Constraints.}}
We define $\varGamma$ as traffic-space constraints, i.e., a set of constraints on malicious functionality and communication protocols, which reflects the requirements of realistic traffic-space objects.
\end{theorem}

\begin{theorem}
\textit{\textbf{Feature-Space Constraints.}}
We define $\varTheta$ as feature-space constraints.
The traffic-space constraints $\varGamma$ determine the feature-space constraints $\varTheta$.
To facilitate the solution, we transform the traffic-space constraints $\varGamma$ into feature-space constraints $\varTheta$.
Given feature object $\mathbf{x} \in \mathcal{X}$, any modification of its feature values can be represented as a perturbation vector $\boldsymbol{\delta} \in \mathbb{R}^{n}$.
If $\boldsymbol{\delta}$ satisfies $\varTheta$, we write $\boldsymbol{\delta} \models \varTheta$.
\end{theorem}

\begin{theorem}
\textit{\textbf{Attack Objective Function.}}
In ML-based NIDSs, $f$ represents ML classifiers that take feature vectors as input and output the malicious probabilities.
Given a feature vector $\mathbf{x} \in \mathcal{X}$, an attack objective function $g$ is defined as follow: $g(\mathbf{x},\hat{\mathbf{x}})=f(\mathbf{x})+\lambda l(\mathbf{x},\hat{\mathbf{x}})$, where $l$ is the distance of malicious feature $\mathbf{x}$ and adversarial feature $\hat{\mathbf{x}}$, and $\lambda$ is a constant weight for $l$.
\end{theorem}

\begin{theorem}
\textit{\textbf{Feature-space Attack.}}
The attacker aims to identify a perturbation vector $\boldsymbol{\delta} \in \mathbb{R}^{n}$ such that $f(\mathbf{x}+\boldsymbol{\delta})<\tau$, where $\tau$ indicates malicious threshold value.
Currently, the attack objective function is $g(\mathbf{x},\boldsymbol{\delta})=f(\mathbf{x})+\lambda l(\mathbf{x},\mathbf{x}+\boldsymbol{\delta})$.
The desired perturbation can be achieved by solving the following optimization problem:
\begin{equation}
    \begin{aligned} \boldsymbol{\delta}^{*}=& \arg \min _{\boldsymbol{\delta} \in \mathbb{R}^{n}} & g(\mathbf{x},\boldsymbol{\delta}) \\ & \text { subject to: } & \boldsymbol{\delta} \models \varTheta \end{aligned} 
\end{equation}

\end{theorem}

\subsubsection{Transfer-based Attacks}
\textcolor{black}{
Transfer-based attacks refer to adversarial attack schemes that are performed by utilizing the transferability of AEs. Formally, let the target classifier with parameters $\theta$, namely $f_{\theta}$, accepts an input $x\in R^{n}$ and makes a prediction $f_{\theta}(x)$. The goal of the adversarial attack can be expressed as generating an AE input $\hat{x}=x+\delta$ to make $f_{\theta}(\hat{x})\neq f_{\theta}(x)$. 
If an AE $\hat{x}$ crafted on $f_{\theta}$ can deceive another model $\phi_{\theta^{\prime}}$ with unknown parameters and architecture, i.e., resulting in $\phi_{\theta^{\prime}}(\hat{x})\neq\phi_{\theta^{\prime}}(x)$, then we say this AE transfers from model $f_{\theta}$ to $\phi_{\theta^{\prime}}$. Typically, the output label sets of the source model $f_{\theta}$ and the target model $\phi_{\theta^{\prime}}$ are consistent. This transferability property means that an adversary can attack against a target black-box model $\phi_{\theta^{\prime}}$ by transferring AEs generating from $f_{\theta}$, which so-called transfer-based attack.
}


\textcolor{black}{
There are three major components affecting the effectiveness of transfer-based attacks: substitute model, substitute dataset, and substitute feature extractor, respectively. Details of the three components are elaborated as follows:
}


\textcolor{black}{
\textit{\textbf{Substitute Model.}} Substitute model refers to a local 
NIDS model that allows adversaries to generate AEs in white-box settings. Such AEs will be transferred to attack the target NIDS. Intuitively, the substitute model is expected to resemble the target model to increase the transferability of the generated AEs. \textbf{The most important part of this paper is the cross-model transferability.}
}

\textcolor{black}{
\textit{\textbf{Substitute Dataset.}} Substitute dataset refers to a dataset used to train the substitute model. The distribution of the substitute dataset is expected to approximate the distribution of the dataset used to train the target model. In this way, the similarity between the substitute model and the target model will increase. However, the substitute datasets that an adversary possesses are usually much smaller than those owned by the MLaaS systems. }

\textcolor{black}{
\textit{\textbf{Substitute Feature Extractor.}} Substitute feature extractor refers to a tool used to extract the features of the substitute dataset. There are two cases regarding the feature extractors mastered by the adversary. In gray-box case, we assume the feature extractor used in the targeted NIDS is known by the attacker. Thus the attacker can build the same extractor and use it to extract features exactly. This may seem extreme, but the features used in ML-based NIDSs are often published in fact. In the more practical black-box case, the attacker has very limited or even no knowledge about the features used in the targeted NIDS. Thus the attacker can only build a substitute extractor based on the domain knowledge. }




\subsubsection{Ours Transfer-based Attacks}
Below we formulate our transfer-based attacks.
Specifically, the attacker first trains a substitute classifier $f'$ locally to approximate the actual ML-based classifier $f$.
The substitute model is an ensemble classifier $f'=(f_i)^k_{ i=1}$ with high transferability, where $f_i$ is the base model, and $k$ is the number of base models.
Second, the attacker selects $m$ non-robust features from the $n$ original features to compose a new feature vector $\mathbf{x}^\circ$.
Third, the attacker evaluates the gradient of the selected feature vector $\mathbf{x}^\circ$ to determine the direction of AEs.
And then, the attacker use remapping function $\mathcal{M}(\hat{x})$ to project $\mathbf{x}^\circ$ back to $\mathbf{x}$ for craft adversarial features $\hat{\mathbf{x}}$ which ensures that $\hat{\mathbf{x}}$ can be transformed into a legal $\hat{\mathbf{t}}$.
In detail, the attacker needs to accomplish two goals: 
\begin{itemize}
\item $\hat{\mathbf{x}}$ must return an erroneous label as benign feature.
\item $\hat{\mathbf{t}}$ (corresponds to the original $\hat{\mathbf{x}})$ must preserve the malicious functionality and communication protocols of the original traffic $\mathbf{t}$.
\end{itemize}
Finally, the adversarial feature $\hat{\mathbf{x}}$ is transferred to the ML-based classifier $f$.

\subsection{Optimizing the Substitute Model}
A critical aspect of transfer-based attacks is the need for a substitute model with high transferability.
The substitute model is optimized by ensemble non-differentiable models and ensemble weight optimization.

\subsubsection{Ensemble Non-differentiable Models}
\cite{liu2016delving} proposed novel ensemble-based approaches to generating AEs and found that if AEs remain adversarial across multiple models, it is likely to transfer to other models.
However, most practical ML-based NIDSs are based on traditional machine learning.
Due to significant differences in the decision boundaries between non-differentiable tree models and differentiable deep learning models, transfer-based attacks using deep learning as a substitute model have a low success rate to attack tree models.
Accordingly, our substitute model is an ensemble of differentiable and non-differentiable models.

\subsubsection{Ensemble Weight Optimization}
A previous study \cite{liu2016delving} assumed equal importance among different models in an ensemble model.
Nevertheless, in practice, the ease of adversarial attack varies significantly among models.
We develop novel weight ensemble approaches and demonstrate that they can craft more transferrable AEs.
Specifically, considering $k$ ML/DL models $(f_i)^k_{ i=1}$, the goal is to find robust adversarial features that can fool all $n$ models simultaneously.
In this case, the objective function $g(\mathbf{x},\boldsymbol{\delta})=\sum_{i=1}^nw_if_i(\mathbf{x})+\lambda l(\mathbf{x},\mathbf{x}+\boldsymbol{\delta})$ signifies the attack loss $g(\mathbf{x},\boldsymbol{\delta},f_i)$ given the natural input $\mathbf{x}$ and the model $f_i$.
Thus, spurred by \cite{wang2019towards}, optimization goal becomes
\begin{equation}
    \underset{\boldsymbol{\delta}}{\operatorname{min}} \underset{\mathbf{w}}{\operatorname{max}} \sum_{i=1}^{n} w_{i} g\left(\mathbf{x},\boldsymbol{\delta},f_{i}\right)-\gamma\|\mathbf{w}-\mathbf{1} / n\|_{2}^{2}
\end{equation}
where $\mathbf{w}$ encodes the difficulty level of attacking each model, and $\gamma $ is a regularization parameter.

\vspace{-0.15in}
\subsection{Important-Sensitive Feature Selection (ISFS)}
According to an interpretation based on cooperative game theory (CoGT) and perturbation, the Important-Sensitive Feature Selection (ISFS) is divided into three key steps.
To begin with, we evaluate the importance of features based on cooperative game theory.
Besides, the sensitive evaluation is achieved by perturbation-based interpretation.
Ultimately, features are classified as robust and non-robust according to the above evaluation results, which means the existence and transferability of adversarial examples (AEs) can be explained.

\subsubsection{The Motivation of ISFS}
For explaining machine learning models, various methods \cite{chang2018explaining,fong2019understanding,fong2017interpretable,lundberg2017unified,covert2020understanding} have been proposed.
Perturbation-based interpretations \cite{fong2019understanding} assess the effect of a single or group features on the model prediction.
However, features generally interact with each other, which means that a single or a group features cannot be assumed as an intrinsically important feature to impact the model prediction significantly. 
A reasonable model should focus more on the overall contribution of features rather than the individual contribution of features.
We claim that perturbation-based interpretations can only be used to determine the model's sensitivity to features, but not their importance.
By taking into account the interactions between features, the CoGT-based interpretation \cite{lundberg2017unified} compensates for the shortcomings of the perturbation-based method.
A feature's importance should be evaluated in terms of its effect on the overall model, which is the average contribution across all the subsets that contain the feature, i.e., its Shapely value.
Our ISFS method combines CoGT-based and perturbation-based interpretations to assess the importance and sensitivity of features.
Based on this, we propose an interpretation of AEs, which is the first step toward understanding the relationship between AEs and explainable AI.
An in-depth understanding of this connection could improve our understanding of AEs and the detector's performance.
We hold the view that the reason for the AEs is that the model learns some independently important and sensitive features, which coincides with the view of \cite{ilyas2019adversarial}.
 
\subsubsection{Feature Importance Evaluation}
\textcolor{black}{
Feature importance evaluation is a critical step in ISFS.
Explainable methods can be used to assess the importance of features.
The Shap \cite{lundberg2017unified} based CoGT for feature evaluation has theoretical guarantees and satisfies additivity.
However, this method is more concerned with the interpretability of individual samples. While the calculation of global importance is only an average of the importance of individual samples. Such global importance evaluation results are unjustified, as features usually interact with each other.
Accordingly, 
\cite{covert2020understanding} derived SAGE (Shapley Additive Global Explanation) by applying the Shapley value to a function representing the predictive power contained in subsets of features.
It is reasonable to assess the importance of features by their predictive power, which can account for feature interactions.
We therefore evaluate the global importance of each feature in the target NIDS by exploiting \cite{covert2020understanding}. A brief description of this scheme is given below.
}



\textcolor{black}{
\textit{\textbf{Theoretical Basis of Shapley.}} 
The Shapley value is used in cooperative game theory to assess the contribution of each player. For this reason, in the interpretable domain, Shapley value can assess the contribution of each feature to the predicted results of the target model, i.e., the importance of each feature. Specifically, we consider multiple features $\mathcal{X}=\{X_1,X_2,...,X_d\}$ play a cooperative game to win a predicted results. The Shapley value $\phi(X_i\mid\mathcal{X})$ measures the importance of the $i$-th feature $X_i$ to the ultimate predicted results contributed by all features in $\mathcal{X}$, as follows:}

\textcolor{black}{
{\scriptsize
\begin{equation}
    \phi(X_i \mid \mathcal{X})= \sum_{X_S \subseteq \mathcal{X} \backslash\{X_i\}} \frac{|S| !(d-|S|-1) !}{d!}(w(X_S\cup\{X_i\})-w(X_S))
\end{equation}}
where $X_S\subseteq \mathcal{X}$ represents all potential subsets of $\mathcal{X}$, $w (\cdot )$ represents the reward function from predicted results. And therefore $w(X_S)$ represents the reward obtained by a set of features $X_S$.}

\textcolor{black}{
\textit{\textbf{Predictive Power of Feature Subsets.}}
To express the feature importance with feature interaction consideration, \cite{covert2020understanding} introduced the concept of predictive power of feature subsets $v_f$. The concept $v_f$ denotes the degradation degree of model prediction when the feature subset $X_S$ is removed from the full feature set $\mathcal{X}$. Let $f_{S}\left(x_{ S}\right)=\mathbb{E}\left[f(X)\mid X_{S}=x_{S}\right]$ represents the average benefit function for a subset $X_S$, then $v_f$ can be formulated as follow: 
\begin{equation}
    v_{f}(X_S)=\underbrace{\mathbb{E}\left[\ell\left(f_{\varnothing}\left(x_{\varnothing}\right), Y\right)\right]}_{\text {Mean prediction }}-\underbrace{\mathbb{E}\left[\ell\left(f_{S}\left(x_{S}\right), Y\right)\right]}_{\text {Using features sets}X_S}
\end{equation}
}


\textcolor{black}{
\textit{\textbf{Importance Evaluation of Each Feature.}} According to the aforementioned Shapley value and predictive power of feature subsets, the optimization objective to evaluate the importance of each feature can be given by:
\begin{equation}
    \min_{\phi_{1},\ldots,\phi_{d}}\;\sum_{X_S\subseteq \mathcal{X}}\frac{d-1}{\binom{d}{|S|}|
S|(d-|S|)}{(}\sum_{i\in S}\phi_{i}-v_{f}(X_S){)}^{2}
\end{equation}
where the Shapley values $\phi_{1},\phi_ {2},\ldots,\phi_{d}$ are the optimization solution, and represent each feature's importance in the model prediction result.
}

\subsubsection{Feature Sensitivity Evaluation}
Next, feature sensitivity (FS) is evaluated for each feature or combination of features.
The evaluation process is straightforward.
We choose a feature $x_i$ or a group of features $(x_1,x_2,...x_o)$ to launch the adversarial attack individually and access the effect of the attack.
For flow-based NIDSs, we calculate the success rate of generated AEs directly.
While we evaluation how much $f$'s performance degrades when the adversarial attack launches in packet-based NIDSs.
Because it is challenging to craft AEs successfully based on single or several features for packet-based models.

\subsubsection{Feature Classification and Selection}
According to feature importance and sensitivity evaluation, we classify features according to the inspiration of \cite{ilyas2019adversarial}.
A robust feature is important, but it is not sensitive, while a non-robust feature is both important and sensitive.
Our ISFS method is to select the top $m$ non-robust features as shown in Algorithm~\ref{algo-fs}.

\begin{algorithm}
	\caption{ISFS}
	\label{algo-fs}
	\begin{algorithmic}[1]
	\Require Full feature vector $\mathbf{x}=(x_0,x_1,...,x_n)$, Number of important features $\tau$, Sensitivity threshold $\epsilon$
	\Ensure Non-robust feature vector $\mathbf{x_\circ}=(x_0,x_1,...,x_m)$
    \State Calculate the FAI values $\varrho(x_i)$ in Algorithm \ref{alg-fai}
    \State Sort by FAI
    \State Select the top $\tau$ FAI
    \State Calculate the FS values $\rho(x_i)$
    \If{$\rho(x_i)>\epsilon$}
    \State $x_i$ is non-robust feature
    \EndIf
    \State Select $m$ non-robust features to compose $x^\circ$
	\State \textbf{return} $x^\circ$
	\end{algorithmic}
\end{algorithm}

\subsubsection{Interpretations}
\label{subsec-interpretation-eta}
There have been a variety of explanations proposed for the phenomenon of AEs, as described in Section~\ref{subsec-reason}.
However, the theory of non-robust features \cite{ilyas2019adversarial} is ideal for NIDSs because the training process is feature-centric.
It is intuitive to explain AEs through non-robust features.
To prove AEs can be directly attributed to the presence of non-robust features, we conduct three sets of experiments on the effect of non-robust features on robustness and accuracy.
Here, we begin with a brief description of our experiment.
The detailed analysis of the interpretation is shown in Section~\ref{subsec-interpretation}.
\begin{itemize}
    \item By gradually removing non-robust features, we can observe how the robustness of the model changes.
    \item To test whether non-robust features can achieve high accuracy, we let models contain only a few non-robust features.
    \item By analyzing the relationship between the similarity of non-robust features and adversarial transferability in two models, we uncover the real reason for the adversarial transferability.
\end{itemize}

\subsection{Gradient Evaluation Based on Zeroth-order Optimization}

It is impossible to apply gradient descent directly in all ML-space when the substitute model is an ensemble model containing tree-based models.
Consequently, we utilize a novel gradient evaluation method based on a Zeroth-order (ZO) optimization as a more generalized gradient solving method.

We aim to estimate the gradient $\triangledown_\mathbf{x}g(\mathbf{x})$ of the ML substitute model $f'$ more accurately to generate adversarial feature $\hat{\mathbf{x}}$.
Different Zeroth-order (ZO) algorithms have been developed \cite{ghadimi2013stochastic,lian2016comprehensive}, and their convergence rates have been rigorously studied under different problem settings.
ZO-SGD (ZO-Stochastic gradient descent) is our gradient evaluation method, which mimics the first-order (FO) process but approximates the gradient through random gradient estimates obtained by comparing function values at random query points.
ZO-SGD gradient estimation is constructed by forward comparing two function values at random directions: 
$\nabla_\mathbf{x}g(\mathbf{x})=(1/ \sigma)[g(\mathbf{x}+\sigma\mathbf{u})-g(\mathbf{x})] \mathbf{u}$, 
where $\mathbf{u}$ is a random vector drawn uniformly from the sphere of a unit ball, and $\sigma> 0$ is a small step size, known as the smoothing parameter. 
The random direction vector $\mathbf{u}$ is drawn from the standard Gaussian distribution.
Since we are evaluating the gradient after feature selection, we use the mask function $\mathfrak{M}$ to keep the randomly generated $\mathbf{u}$ restricted to a specific feature location.
In the end, we evaluate the gradient with a population of $q$ points sampled under this scheme: $\nabla \mathbb{E}[g(\mathbf{x})] \approx \frac{1}{\sigma q} \sum_{i=1}^{q} \mathbf{u}_{i} [g\left(\mathbf{x}+\sigma \mathbf{u}_{i}\right)-g(\mathbf{x})]$.
The detailed steps of zero-order optimization are shown in Algorithm \ref{algo-gra-eva}.

\begin{algorithm}
	\caption{Gradient Evaluation Method}
	\label{algo-gra-eva}
	\begin{algorithmic}[1]
	\Require Objective function $g(\mathbf{x})$ for input $\mathbf{x}$, Search variance $\sigma$, Number of samples $q$, Feature dimensionality $n$, Variance of Gaussian distribution $\zeta$
	\Ensure Estimate of $\nabla_\mathbf{x} g(\mathbf{x})$
	\State $\nabla_{tmp}\leftarrow \textbf{0}_n$
	\For{$i=1$ \textbf{to} $q$}
    \State Randomly sample Gaussian vectors $\mathbf{u}_i\leftarrow \mathcal{N}(0,\zeta^2)$
    \State $\mathbf{u}_i=\mathfrak{M}({\mathbf{u}_i})$
	\State $\nabla_{tmp}\leftarrow \nabla_{tmp}+ [g(\mathbf{x}+\sigma\cdot\mathbf{u}_i)-g(\mathbf{x})]\cdot\mathbf{u}_i$
	\EndFor
	\State \textbf{return} $\nabla_{tmp}/q\sigma$
	\end{algorithmic}
\end{algorithm}

\vspace{-0.15in}
\subsection{Crafting Adversarial Examples in Domain Constraints}
We craft the adversarial examples (AEs) in domain constraints by two steps: (1) Multi-step min-max iterative optimization, (2) Projecting AEs in domain constraints.
\subsubsection{Multi-step Min-max Iterative Optimization}
\label{subsec-multi-step}
Next, we consider crafting adversarial examples (AEs) as a multi-step min-max optimization problem.
In multi-step optimization, the typical search direction for AEs in feature-space is based on gradient evaluation of objective functions.
The detailed steps of multi-step optimization are shown in Algorithm~\ref{algo-mul}.
The $\eta$ is the number of interactions. 
At min-step iteration, we adopt $l_2$-norm to calculate the distance $l$ of malicious feature vector $\mathbf{x}$ and adversarial feature vector $\hat{\mathbf{x}}$.
And we craft AEs as $\mathbf{x}^t=\mathbf{x}+\varepsilon \operatorname{sgn}\left(\nabla_{\mathbf{x}} g(\mathbf{x},\hat{\mathbf{x}})\right)$ in this iteration, then project $\hat{\mathbf{x}}$ onto the boundary of the feasible region $\varTheta$.
The details of the feasible region are shown in Section~\ref{subsec-project}.
At max-step iteration, we update the weights of the ensemble substitute model as $w_{t+1}=w_t+\nabla_{w}g(\mathbf{\mathbf{x}},\hat{\mathbf{\mathbf{x}}})$.
Finally, we stop the loop when $\eta$ exceeds the maximum number of iterations or has successfully crafted AEs.

\begin{algorithm}
	\caption{Multi-step min-max iterative optimization}
	\label{algo-mul}
	\begin{algorithmic}[1]
	\Require Malicious feature vectors $\mathbf{x}$
	\Ensure Adversarial feature vectors $\hat{\mathbf{x}}$
	\For{$t=1$ \textbf{to} $\eta$}
    \State min-step:
    \State $\mathbf{x}^{t+1}=\mathbf{x}^{t}+\varepsilon \operatorname{sgn}\left(\nabla_{\mathbf{x}} g(\mathbf{x},\hat{\mathbf{x}})\right)$
    \State $\mathbf{x}^{t+1}=\mathcal{M}(\mathbf{x}^{t+1})$
    \State max-step:
    \State $w_{t+1}=w_t+\nabla_{w}g(\mathbf{\mathbf{x}},\hat{\mathbf{\mathbf{x}}})$
    \If{$\mathbf{x}^{t+1}$ success}
    \State break
    \EndIf
    \EndFor
    \State \textbf{return} $\hat{\mathbf{x}}$
	\end{algorithmic}
\end{algorithm}

\textcolor{black}{
\subsubsection{Projecting Adversarial Examples in Traffic-space Constraints}
\label{subsec-project}
We consider how to make feature modifications satisfy the traffic-space constraints.
As stated in Section~\ref{subsec-multi-step}, we solve the constrained optimization problem in a multi-step iterative manner, and each step projects the result into a feasible region.
Here we describe how to ensure that the generated adversarial features conform to the traffic-space constraints.
}

\textcolor{black}{
Specifically, unlike in the image recognition domain where AEs can modify each pixel value individually, applying AEs in network traffic is much more challenging because there are considerable constraints. 
For example, there are some features of network traffic that cannot be modified and some features can only be modified within valid ranges when generating AEs. If feature values are modified to out-of-bounds values, the attack will invalidate because the generated traffic loses its attack semantics and cannot be implemented in the real world. 
We detail the inherent constraints on network traffic from the following four aspects:
\begin{itemize}
    \item \textit{\textbf{Attack types}}: 
    AEs must retain the attack function of the original traffic. The attack type specifies a set of traffic characteristic values that must be kept constant. That is, such feature values cannot be modified to preserve the underlying semantics of attack.
    \item \textit{\textbf{Protocol types}}: 
    AEs must preserve the network protocol correctness to ensure that the AEs will not be discarded.
    Protocol types specify that certain subsets of characteristics must be zero, as certain characteristics are protocol specific (e.g. TCP flags). In addition, flags and services allow a protocol to have only certain characteristic values.
    \item \textit{\textbf{Correlation between features}}: There are some features whose values depend on other features in network traffic. For example, the mean of packet payloads in the forward direction can be calculated based on two other features, the total length of payloads in the forward direction and the total number of forward packets. AEs are not allowed to modify the correlation between these features.
    \item \textit{\textbf{Valid range}}: 
    For each attack type, the values of features are only taken within a certain range. Only the values within this range are valid values. AEs should avoid modifying feature values to out-of-bounds.
\end{itemize}
}






\textcolor{black}{
We define three ways to make the generated adversarial examples satisfy the traffic-space constraint.}

\textcolor{black}{
\textbf{Mask}: We set features that are likely to break network protocols and potentially corrupt malicious functions to be unmodifiable using the mask function.}

\textcolor{black}{
\textbf{Clip}:
For each type of attack, we find the minimum and the maximum value of each feature and define the range between the minimum and maximum as the valid range.
Then, after applying the perturbation to original attack traffic during the process of crafting adversarial examples, we enforce the valid range for each feature of attack traffic by clipping. 
For example, if the feature value of $x_1$ of a DoS attack traffic becomes $x_{1}+\Delta x_{1}<d_{\min }$ minafter a perturbation is applied, then we clip the value and replace it with $d_{min}$.}

\textcolor{black}{
\textbf{Equation constraints: }
It is unrealistic to assume that all features are independent of each other. Adding perturbations to related features independently may compromise the validity of the instances. We therefore make consistent modifications to interrelated features, and we use equation constraints to ensure consistency of feature associations.}

\textcolor{black}{
To enforce the above domain constraints, we use the remapping function $\mathcal{M}$ to adjust the perturbed features.
For example, when a feature can be calculated based on other features, the remapping function $\mathcal{M}$ directly utilizes the calculated value rather than being updated by the gradient.
Therefore, the function of $\mathcal{M}$ can make the AEs on feature vectors $\mathbf{x}$ satisfy the above traffic-space constraints.}

\section{experiment Evaluation}
\label{sec-evaluation}
In this section, we first empirically evaluate the effectiveness of ETA with four baselines.
Second, the ablation experiments demonstrate that our substitute model optimization and feature selection algorithms are effective.
And then, we conduct three interpretation experiments to verify the correctness of our interpretation.
We finally test the sensitivity of hyper-parameters.
Overall, our experiments cover the following aspects:
\begin{itemize}
    \item \textbf{Experimental Setup.} We introduce implementation and setup of ML-based NIDSs systems, datasets, evaluation metrics, and baseline attacks in this study (\ref{subsec-exp-setting}).
    \item \textbf{Adversarial Attack Effectiveness.} We demonstrate the effectiveness of ETA by comparing the effects of different classifiers and datasets with baseline attacks (\ref{subsec-adv-effect}).
    \item \textbf{Ablation Study.} Using ablation studies, we assess the impact of each component of our ETA system (\ref{subsec-ablation}).
    \item \textbf{Interpretation.} The experiment explains the reasons for AEs and adversarial transferability in NIDSs by removing, using, and comparing non-robust features (\ref{subsec-interpretation}). Additionally, we uncover two major misconceptions regarding machine learning in NIDS systems (~\ref{subsec-use-non-robust} and~\ref{subsec-understand-case}).
    \item \textbf{Hyper-parameters Sensitivity.} We evaluate six hyper-parameters in ETA (\ref{subsec-hyper}).
    \item \textbf{Evaluations in Real Environment.}
    We evaluate the ETA in real environment. (\ref{subsec-real}).
\end{itemize}

\subsection{Experimental Settings}
\label{subsec-exp-setting}
Here we discuss the settings of our experiments.

\subsubsection{Hardware $\&$ Software}
We first provide statements of hardware and software in our study.

\begin{itemize}
    \item \noindent\textbf{Hardware.} Our experiments are evaluated on a MacBook Pro using 64-bit MacOS system with Intel Core i7-9700K 3.6 GHz 8-Core Processor, 16 GB physical memory, 256GB SSD.
    \item \noindent\textbf{Software.} The software implementation of ETA is based on Python 3 with several packages, including scikit-learn for ML models and pytorch for DL models.
\end{itemize}

\subsubsection{Dataset}
Many public, labeled data sets of network traffic are available.
Two of these datasets are: \textit{\textbf{CIC-IDS2017}} and \textit{\textbf{Kitsune}}.
CIC-IDS2017 \cite{sharafaldin2018toward} collects traffic for common attacks in a large-scale testbed, covering many common devices and middleboxes in the network.
Kitsune\cite{mirsky2018kitsune} is lightweight traffic data in video IoT networks.
We select eight sub-datasets from two datasets as shown in Table~\ref{table-datasets}.
Meanwhile, we divide the sampled data into three parts according to 3:1:1 for model training, verification and testing respectively.
The data for training and testing are shown in the Table~\ref{table-datasets}, and each part contains both benign (B) and malicious (M) samples.
\begin{table}[h]
    \caption{The details of datasets used in experiments.}
    \label{table-datasets}
    \begin{tabular}{@{}llllll@{}}
    \toprule
    \multirow{2}{*}{Feature type} & \multirow{2}{*}{Datesets} & \multicolumn{2}{l}{Training Phase} & \multicolumn{2}{l}{Test Phase} \\ \cmidrule(l){3-6} 
                                  &                           & B      & M         & B    & M     \\ \cmidrule(r){1-6}
    \multirow{4}{*}{Flow-based}   & IDS17-Botnet               &8724    & 1180      &2908   &393                 \\
                                  & IDS17-DDoS              &91904           &25101      &43968  &8366                \\
                                  & IDS17-Bruteforce        &25908         &9205         &8636         &3068                \\
                                  & IDS17-WebAttack         &10083          &1266          &3361     &422                 \\
    \multirow{4}{*}{Packet-based} & Kitsune-Mirai           &32838               &28104              &10946         &9638                 \\
                                  & Kitsune-SYN-DoS     &13821               &6968              &4607         &2323                 \\
                                  & Kitsune-Fuzzing           &73020              &64920              &24340         &21640                 \\
                                  & Kitsune-SSDP-Flood             &23739               &21594               & 7913       &7198                 \\ 
    \bottomrule
    \end{tabular}
    \vspace{-0.2in}
    \end{table}

\begin{table*}[]
    \caption{The attack effectiveness on NIDSs with feature extractors and ML classifiers}
    \label{table-attack_effect} 
  \begin{center}
  \begin{subtable}{.49\textwidth}
  \centering
      \caption{CicFlowMeter}
      \label{table-attack_effect_cic} 
      \resizebox{1\textwidth}{!}{
      \begin{tabular}{@{}lllllllll@{}}
          \toprule
          \multirow{2}{*}{Model} &                        &                       & \multicolumn{6}{c}{Adversarial Attack (ASR)--higher is better}                                                                       \\ 
          \cmidrule(l){2-9} 
                                 & \multicolumn{1}{c}{F1} & \multicolumn{1}{c}{R} & \multicolumn{1}{c}{JSMA} & \multicolumn{1}{c}{C\&W} & \multicolumn{1}{c}{ZOO} & \multicolumn{1}{c}{HSJA} & \multicolumn{1}{c}{GAN} & \multicolumn{1}{c}{Ours} \\ \midrule
          MLP                 & 0.96              & 0.98             & 98.05\%               & 74.19\%             & 53.43\%                & 99\%                 & 40.12\%                   & \textbf{98.09\%}    \\
          AlertNet            & 0.89              & 0.99             & 15\%                  & 0.12 \%             & 0.51\%                 & 0\%                  & 42.34\%                   & \textbf{74.02\%}    \\
          IDSNet              & 0.82              & 0.93             & 36.05\%               & 2.07\%              & 28.01\%                & 49.00\%              & 39.32\%                   & \textbf{75.03\%}    \\
          DeepNet             & 0.91              & 0.98             & 4.79\%                & 6.09\%              & 24.01\%                & 4.50\%               & 43.12\%                   & \textbf{77.89\%}    \\
          RF                  & 0.98              & 0.99             & 64.85\%               & 47.08\%             & 100\%                  & 100\%                & 75.23\%                   & \textbf{98.45\%}    \\
          Xgboost             & 0.99              & 0.99             & 35.46\%               & 45.78\%             & 94.03\%                & 69\%                 & 69.54\%                   & \textbf{97.45\%}    \\
          MaMPF               & 0.89              & 0.98             & 12\%                  & 1.29\%              & 1.42\%                 & 11\%                 & 42.45\%                   & \textbf{73.34\%}     \\
          FS-Net              & 0.89              & 0.99             & 13\%                  & 0.12\%              & 0.90\%                 & 23\%                 & 46.78\%                   & \textbf{76.38\%}    \\ 
          Kitnet              & 0.82              & 0.96             & 8.33\%                & 19.05\%             & 23.15\%                & 25.23\%              & 21.34\%                   & \textbf{68.10\%}     \\
          Diff-RF             & 0.99              & 0.98             & 43.81\%               & 1.19\%              & 35.56\%                & 35.56\%              & 25.45\%                   & \textbf{78.57\%}     \\
          \bottomrule
          \end{tabular} }
  \end{subtable}
  \hfill
  \begin{subtable}{.49\textwidth}
  \centering
      \caption{AfterImage}
      \label{table-attack_effect_after}
      \resizebox{1\textwidth}{!}{
      \begin{tabular}{@{}lllllllll@{}}
          \toprule
          \multirow{2}{*}{Model} &                        &                       & \multicolumn{6}{c}{Adversarial Attack (ASR)--higher is better}                                                                       \\ 
          \cmidrule(l){2-9} 
                                 & \multicolumn{1}{c}{F1} & \multicolumn{1}{c}{R} & \multicolumn{1}{c}{JSMA} & \multicolumn{1}{c}{C\&W} & \multicolumn{1}{c}{ZOO} & \multicolumn{1}{c}{HSJA} & \multicolumn{1}{c}{GAN} & \multicolumn{1}{c}{Ours} \\ \midrule
          MLP                 & 0.94              & 0.93             & 89.97\%               & 98.72\%             & 67.79\%                & 51\%                 & 47.34\%                   & \textbf{84.62\%}           \\
          AlertNet            & 0.91 & 0.91 & 88\% & 97.23\% & 56.32\% & 52\% & 56.23\% & \textbf{83.34}\%     \\
          IDSNet              & 0.9 & 0.91 & 86.23\% & 96.24\% & 65.20\% & 56.12\% & 54.34\% & \textbf{82.65}\%     \\
          DeepNet             & 0.92 & 0.93 & 83.35\% & 88.34\% & 18.27\% & 45.56\% & 57.34\% & \textbf{85.56}\%     \\
          RF                  & 0.99 & 0.99 & 0.00\% & 12.10\% & 14\% & 12\% & 26.32\% & \textbf{62.76}\%     \\
          Xgboost             & 0.99 & 0.99 & 7.35\% & 18.12\% & 10.27\% & 10\% & 28.56\% & \textbf{59.40}\%     \\
          MaMPF               & 0.9 & 0.85 & 73\% & 87.34\% & 56.30\% & 46\% & 67.34\% & \textbf{82.10}\%     \\
          FS-Net              & 0.9 & 0.84 & 72\% & 85.34\% & 55.60\% & 43\% & 65.43\% & \textbf{83.20}\%     \\ 
          Kitnet              & 0.94 & 0.94 & 1.21\% & 2.56\% & 2.56\% & 32.22\% & 34.56\% & \textbf{69.66}\%      \\
          Diff-RF             & 0.9 & 0.88 & 4.56\% & 45.01\% & 25.23\% & 42.39\% & 46.32\% & \textbf{56.84}\%      \\
          \bottomrule
          \end{tabular}}
  \end{subtable}
  \end{center}
  \vspace{-0.15in}
  \end{table*}
\begin{figure*}
    \centering
    \begin{subfigure}{.49\textwidth}
     \includegraphics[width=\textwidth]{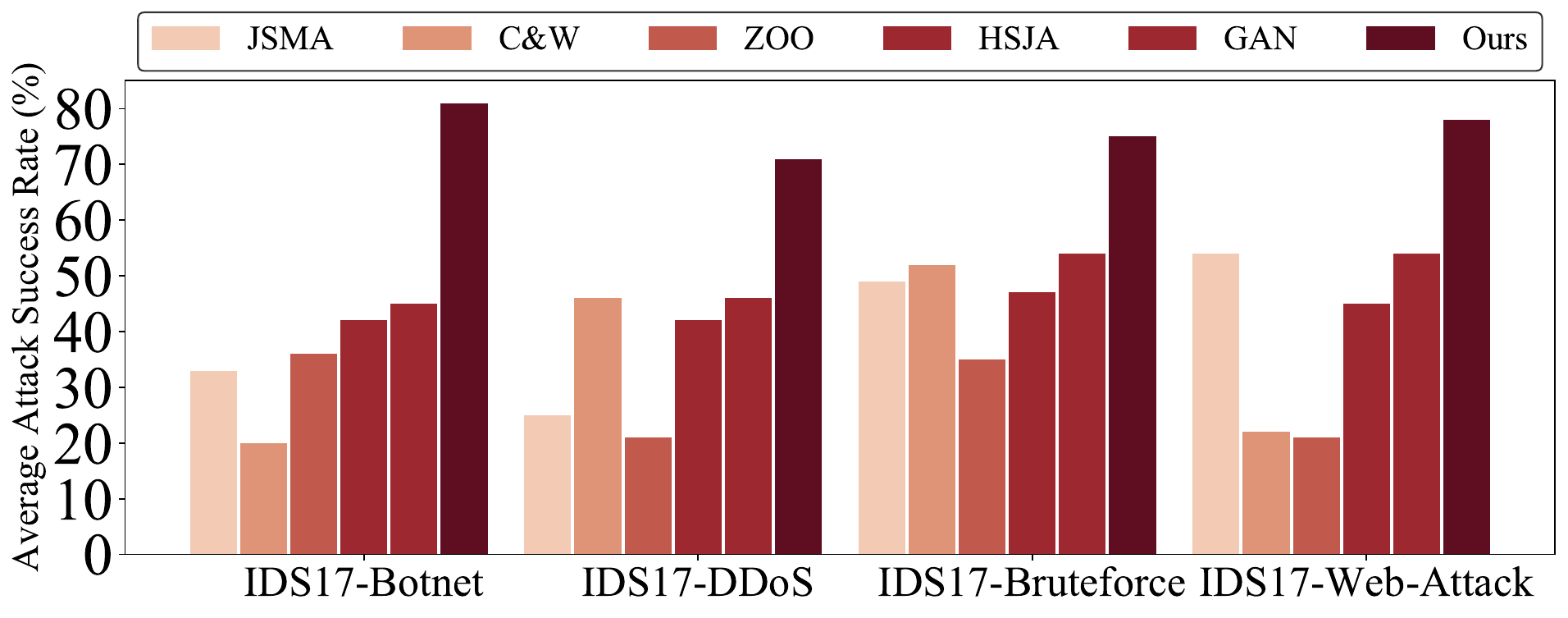}
    \end{subfigure}
    \hfill
    \begin{subfigure}{.49\textwidth}
     \includegraphics[width=\textwidth]{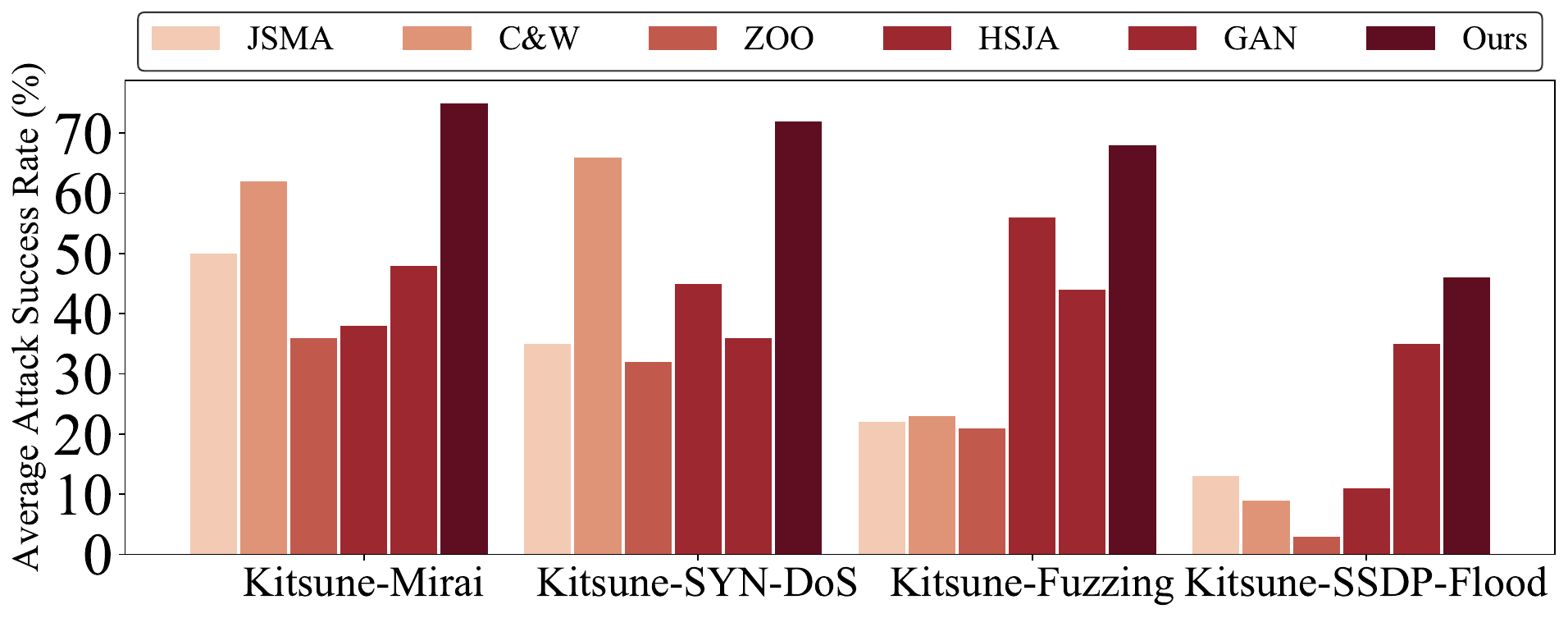}
   \end{subfigure}
   \caption{Adversarial attack effectiveness of our attacks compared with baselines on different datasets (\textit{higher is better}).}
   \label{fig-attack-eff-datasets}
   \vspace{-0.06in}
  \end{figure*}

\subsubsection{Target NIDSs}
Regarding the target NIDSs, we describe their feature extractors for extracting input traffic features and the machine learning models for classification and detection.

\textcolor{black}{
\textbf{Feature Extractors.} We extract packet-level and flow-level features of network traffic, respectively. In particular, temporal features such as packet length sequence or packet arrival interval sequence of flows are also leveraged as the input of the classifiers. In our experiments, \textit{\textbf{CICFlowMeter}}\cite{sharafaldin2018toward} is used to extract flow-level features in CIC-IDS2017, which can extract several statistics (e.g., size, count, and duration) of connections. \textit{\textbf{AfterImage}}\cite{mirsky2018kitsune} is used to extract packet-level features in Kitsune, which computes incremental statistics of packet size, count, and jitter in various damped time windows.
\textit{\textbf{Temporal feature}}: Temporal features such as packet length sequence or packet arrival interval sequence of flows are leveraged as the input of the NIDSs classifiers.
}

\textbf{ML classifiers.} 

\emph{Statistical feature-based methods.}
To test the transferability of different models, the ETA is evaluated on several kinds of NIDSs, including classical machine learning and deep learning. 
Specifically, we explore differentiable gradient models.
A few researchers began employing simple deep learning models to detect malicious traffic. 
Such as, AlertNet \cite{vinayakumar2019deep}, DeepNet \cite{gao2020malicious}, and IdsNet\cite{zolbayar2022generating} are based on fully connected perceptrons with ReLU activation functions, batch normalization to improve training performance, and dropout to prevent overfitting.
For non-differentiable models, we primarily use the decision trees (DT), Random forests (RF), XGboost (Xgb.), and the soft ensemble model, which assigns different weights to different differentiable and non-differentiation models (Ens.).

\textcolor{black}{
\emph{Temporal feature-based methods.}
It is common to apply a sequence model to traffic classification, and this method can also be used to detect malicious traffic as well.
According to the sequence of packet lengths and packet arrival intervals, flows are classified using time series models such as Markov Models and Recurrent Neural Networks.
Liu et al. \cite{liu2018mampf} proposed the Multi-attribute Markov Probability Fingerprints (MaMPF), for encrypted traffic classification ,which can captures the time-series packet lengths effectively using power-law distributions and relative occurrence probabilities of all considered applications. 
Liu et al. \cite{liu2019fs} applied the recurrent neural network to the encrypted traffic classification problem and propose the Flow Sequence Network (FS-Net) which is an end-to-end classification model that learns representative features from the raw flows.
}

\emph{Anomaly Detection methods.} To further explore the attack performance of our ETA system against the latest NIDSs, we select two state-of-the-art anomaly detection models.

\begin{itemize}
    \item KitNET \cite{mirsky2018kitsune} is an online unsupervised anomaly detector powered by an ensemble of auto-encoders that identifies anomalies using reconstruction errors.
    \item The Diff-RF \cite{marteau2021random} is an ensemble approach based on random partitioning binary trees that is capable of detecting point-wise and collective anomalies.
\end{itemize}

\subsubsection{Baseline Attacks}
\label{subsec-baseline}
We evaluate the ETA system with several baseline feature-space attacks as follows:
\begin{itemize}
    \item JSMA: \cite{qiu2020adversarial} first utilized the model extraction technique to replicate the target model and then use saliency maps to disturb the critical features for crafting AEs.
    \item C\&W:  As the only transfer-based attack in NIDSs, \cite{yang2018adversarial} first trained a DL-based substitute model, and then use C\&W to generate AEs.
    \item ZOO: The zeroth-order optimization (ZOO) attack was presented in Chen et al.\cite{chen2017zoo}.  For fairness, we set the query model of the ZOO to an ensemble model.
    \item HSJA: The HSJA (Hopskipjumpattack) method \cite{chen2020hopskipjumpattack} combines decision-based and zero-order optimization. It is the current best decision-based method. We improve it as a transfer-based attack method using the ensemble model as the substitute model.
    \textcolor{black}{\item GAN: Nasr et al. \cite{nasr2021defeating} apply generative adversarial networks (GAN) to generate blind adversarial perturbations that follow the statistical distributions expected from the target protocol.}
\end{itemize}

\subsubsection{Evaluation Metrics}
Metrics used in this work can be classified into two categories:
\begin{itemize}
    \item \noindent\textbf{Detection Performance}: For performance evaluation, we use typical machine learning metrics, including Precision (P), Recall (R), F1-score (F1) and ROC/AUC (AUC).
    \item \noindent\textbf{Attack Effectiveness} ($ASR$ and $ASR_{avg}$): Specifically, $ASR$ measures the percentage of AEs that have already evaded the NIDSs. 
    To test the transferability of the attack, we also define a metric to evaluate the average attack success rate ($ASR_{avg}$) against all models.
\end{itemize}

\subsection{Effectiveness of Adversarial Attack}
\label{subsec-adv-effect}

\textcolor{black}{We first evaluate the effectiveness ($ASR$) of ETA with five baselines (JSMA, C\&W, ZOO, HSJA, GAN) in two traffic sets (IDS17-Botnet, Kitsune-Mirai), which are flow-based and packet-based, respectively. }
Furthermore, we compare the average attack success rate ($ASR_{avg}$) of our attacks and baselines on eight different sub-datasets.

\begin{table*}[]
\caption{Effectiveness of Adversarial Attack with Limited Knowledge.}
\begin{subtable}{.49\textwidth}
     \caption{Limited knowledge of the features (ASR (\%))}  
     \label{tab_knowledge_fea}
     \centering
     \begin{tabular}{@{}llllll@{}}
        \toprule
        Knowledge & MLP & AlertNet & RF & XGBoost & FS-Net \\ 
        \midrule
        10\%  & 45\%    & 55\%    & 65\%    & 66\%    & 35\%    \\
        20\%  & 55\%    & 67\%    & 77\%    & 78\%    & 47\%    \\
        40\%  & 76.85\% & 86.85\% & 99.85\% & 86.85\% & 56.85\%  \\
        80\%  & 99.12\% & 98.22\% & 99.85\% & 99.42\% & 89.12\% \\
        100\% & 99.12\% & 98.22\% & 99.85\% & 99.42\% & 96.12\% \\
        \bottomrule
        \end{tabular}
\end{subtable}
\begin{subtable}{.49\textwidth}
\caption{Limited knowledge of the datasets (ASR (\%))} 
\label{tab_knowledge_dataset}
     \begin{tabular}{@{}llllll@{}}
        \toprule
        Knowledge & MLP & AlertNet & RF & XGBoost & FS-Net \\ 
        \midrule
        1\%   & 96\%  & 92\%  & 91\%  & 90\%  & 91\% \\ 
        2\%   & 97\%  & 96\%  & 97\%  & 96\%  & 92\%  \\
        4\%   & 98\%  & 98\%  & 98\%  & 99\%  & 99\%  \\
        8\%   & 100\% & 100\% & 100\% & 100\% & 100\% \\
        16\%  & 100\% & 100\% & 100\% & 100\% & 100\% \\
        32\%  & 100\% & 100\% & 100\% & 100\% & 100\% \\
        64\%  & 100\% & 100\% & 100\% & 100\% & 100\% \\
        100\% & 100\% & 100\% & 100\% & 100\% & 100\% \\
        \bottomrule
        \end{tabular}
\end{subtable}
\end{table*}

\subsubsection{Attack Effectiveness Comparison between Different Classifiers}
As shown in Table~\ref{table-attack_effect}, the four baselines (JSMA, C\&W, ZOO, HSJA) prefer to attack a specific model.
However, it performs poorly on other models, indicating that it is brutal to transfer to other models.
\textcolor{black}{For example, JSMA achieves a high ASR of 98\% on MLP but shows weak attack performance on DeepNet and Kitnet, with success rates of only 4.79\% and 8.33
\%, respectively (see Table \ref{table-attack_effect_cic}).} This discrepancy can be attributed to significant differences in decision boundaries between the two models (MLP and Kitnet). 
\textcolor{black}{
On average, GAN-based methods can achieve moderate detection rates, but they are not significantly higher compared to our methods.}
Table~\ref{table-attack_effect_after} shows almost identical results.
It also can be observed that the MLP-based surrogate model achieves a high ASR of 89.9\% using the JSMA attack on the MLP model itself. However, the attack effectiveness is considerably weaker on RF and Xgboost, with success rates of only 0\% and 7.35\% respectively (refer to Table \ref{table-attack_effect_cic}).
This is because the decision boundary varies significantly between the two models (e.g., MLP and RF), and adversarial examples (AEs) cannot easily be transferred to other models if they only attack against local models.
By contrast, our approach performs relatively evenly on each model, achieving an average attack success rate $(ASR_{avg})$ over 70\%.
Moreover, among all the methods, only our method is effective for the latest anomaly detection model (Kitnet, Diff-RF), achieving a $ASR$ of over 55\%.
Obviously, experimental results show that our method (ETA) can achieve excellent attack transferability.


\subsubsection{Attack Effectiveness Comparison between Different Datasets}
Fig.~\ref{fig-attack-eff-datasets} presents the $ASR_{avg}$ of our attacks and baselines on eight sub-datasets.
\textcolor{black}{In five baselines, GAN aeshchieves the best performance, leading to around 46\% $ASR_{avg}$ on eight sub-datasets.}
However, ZOO performs poorly, only achieving around 28\% $ASR_{avg}$ in all datasets.
Moreover, the results shows that our method (ETA) demonstrates extremely superior performance with around 76\% $ASR_{avg}$ in Dataset IDS17, which largely outperforms 30\% (baselines).
The reason is that our method optimizes the substitute model and pinpoints the root cause of adversarial transferability.

\subsection{Effectiveness of Adversarial Attack with Limited Knowledge}

\subsubsection{Limited knowledge of the features}

\textcolor{black}{We have evaluated the effectiveness of our attack with a feature extractor that is completely transparent to the attacker as shown in Table.~\ref{tab_knowledge_fea}.
Our attack now extends to features used by targeted NIDS with limited knowledge. 
In particular, we evaluate five types of attackers that know the 100\%, 80\%, 40\%, 20\% and 0\% features that are accurate for NIDSs.
The only difference is the surrogate feature extractor used by the attacker.
The experimental results demonstrate that when the attacker has more than 40\% of the features, using these features to train the feature extractor can also have a good attack effect on the target model. Due to the fact that our method attacks the target model by selecting non-robust features, the number of features selected is small, and the attack is still successful as long as the target model uses these non-robust features.}

\subsubsection{Limited knowledge of the datasets}
\textcolor{black}{
We next evaluate the effect of the substitute dataset on the effectiveness of the attack as shown in Table.~\ref{tab_knowledge_dataset}.
The substitute models use the target model.
We select a random portion of the dataset from the original training set to evaluate the success rate of the attack. 
The experimental results show that even if the substitute dataset is only 1\% of the original training set, a high attack success rate can still be achieved.
The results show that attacks with high success rates can be achieved even if the target dataset is not available.}

\subsection{Ablation Study}
\label{subsec-ablation}

\begin{figure*}
  \centering
 \begin{subfigure}{.26\textwidth}
  \includegraphics[width=\textwidth]{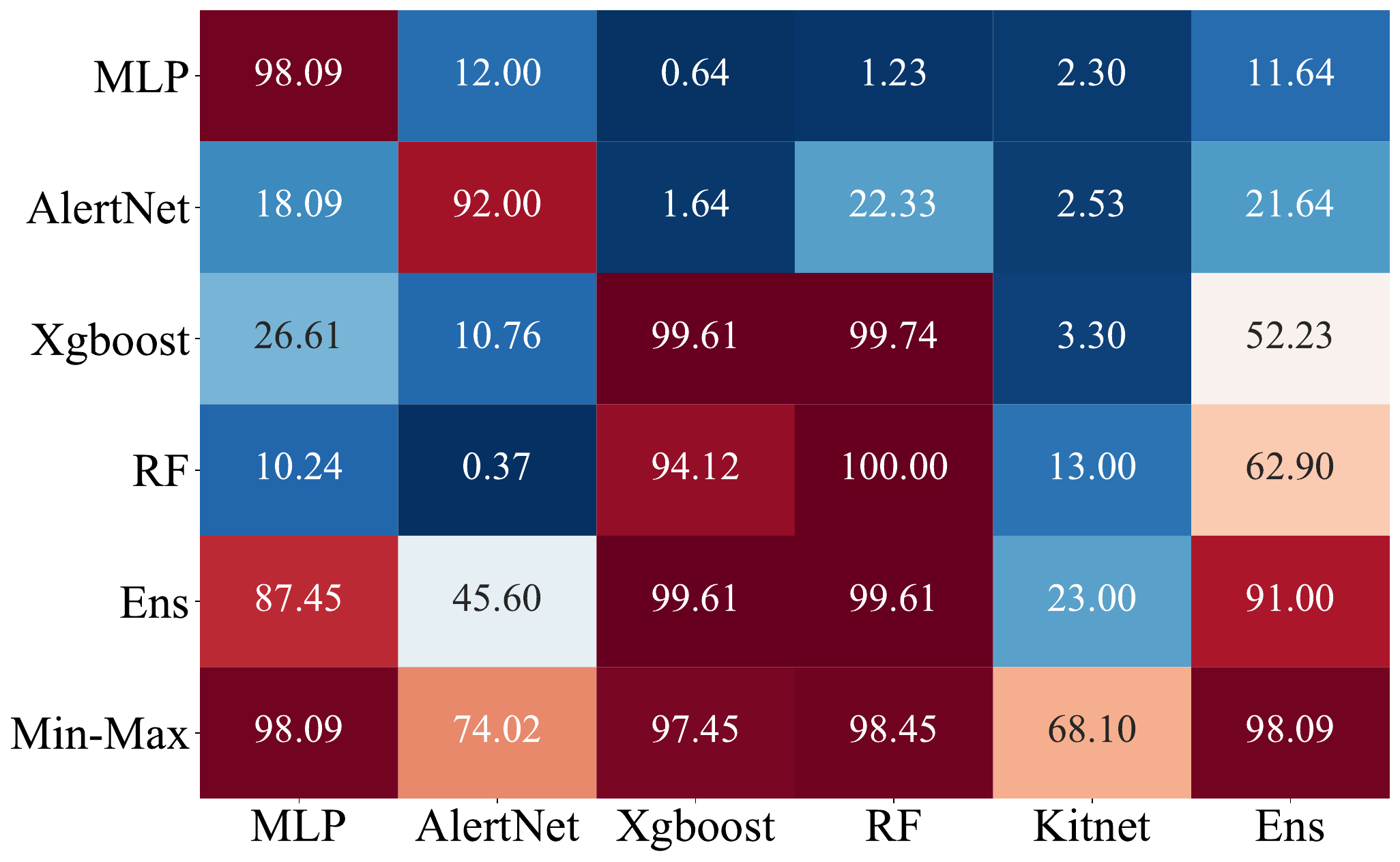}
  \caption{IDS17-Botnet}
 \end{subfigure}
 \hfill
 \begin{subfigure}{.22\textwidth}
  \includegraphics[width=\textwidth]{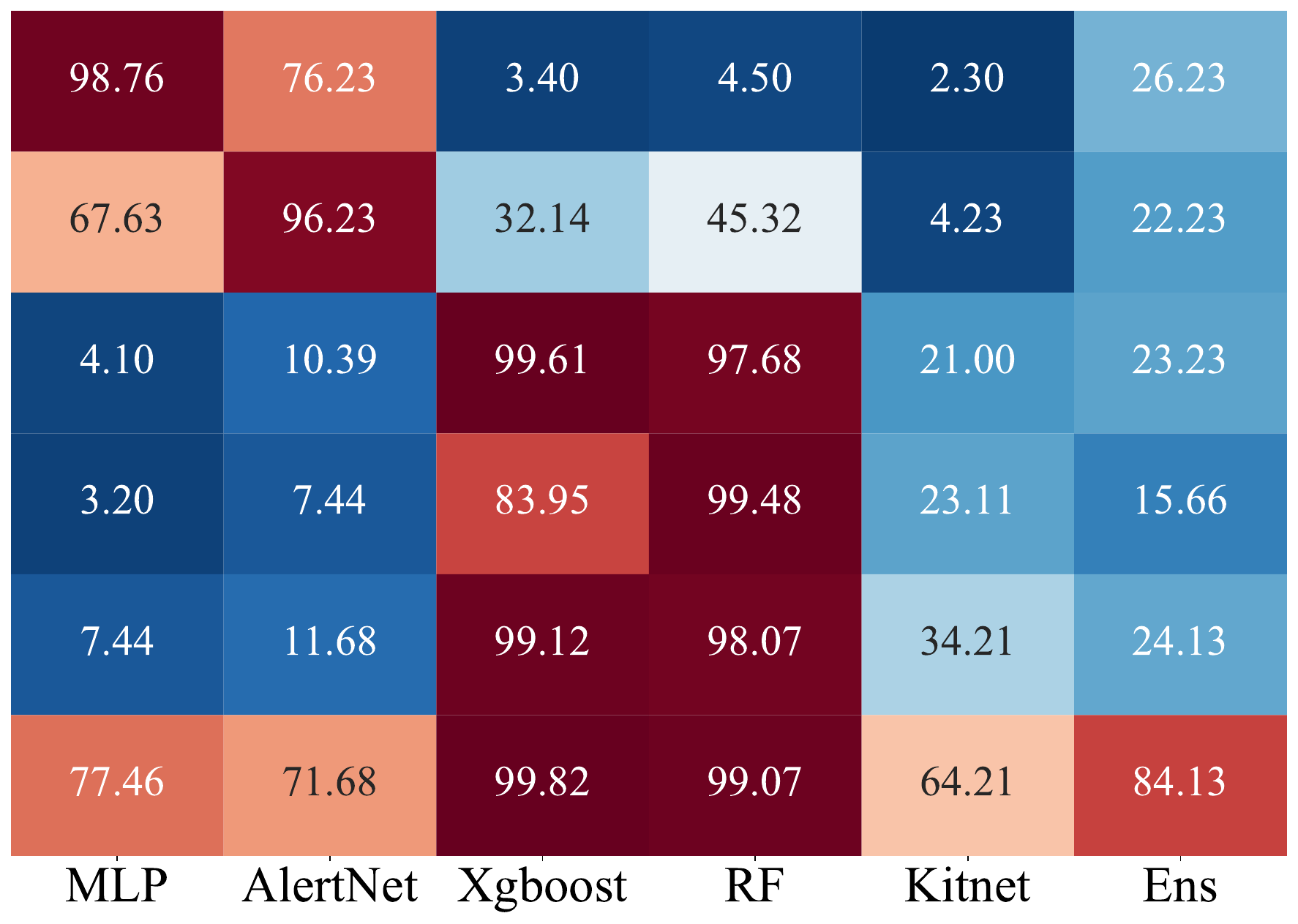}
  \caption{IDS17-DDoS}
 \end{subfigure}
 \hfill
 \begin{subfigure}{.22\textwidth}
  \includegraphics[width=\textwidth]{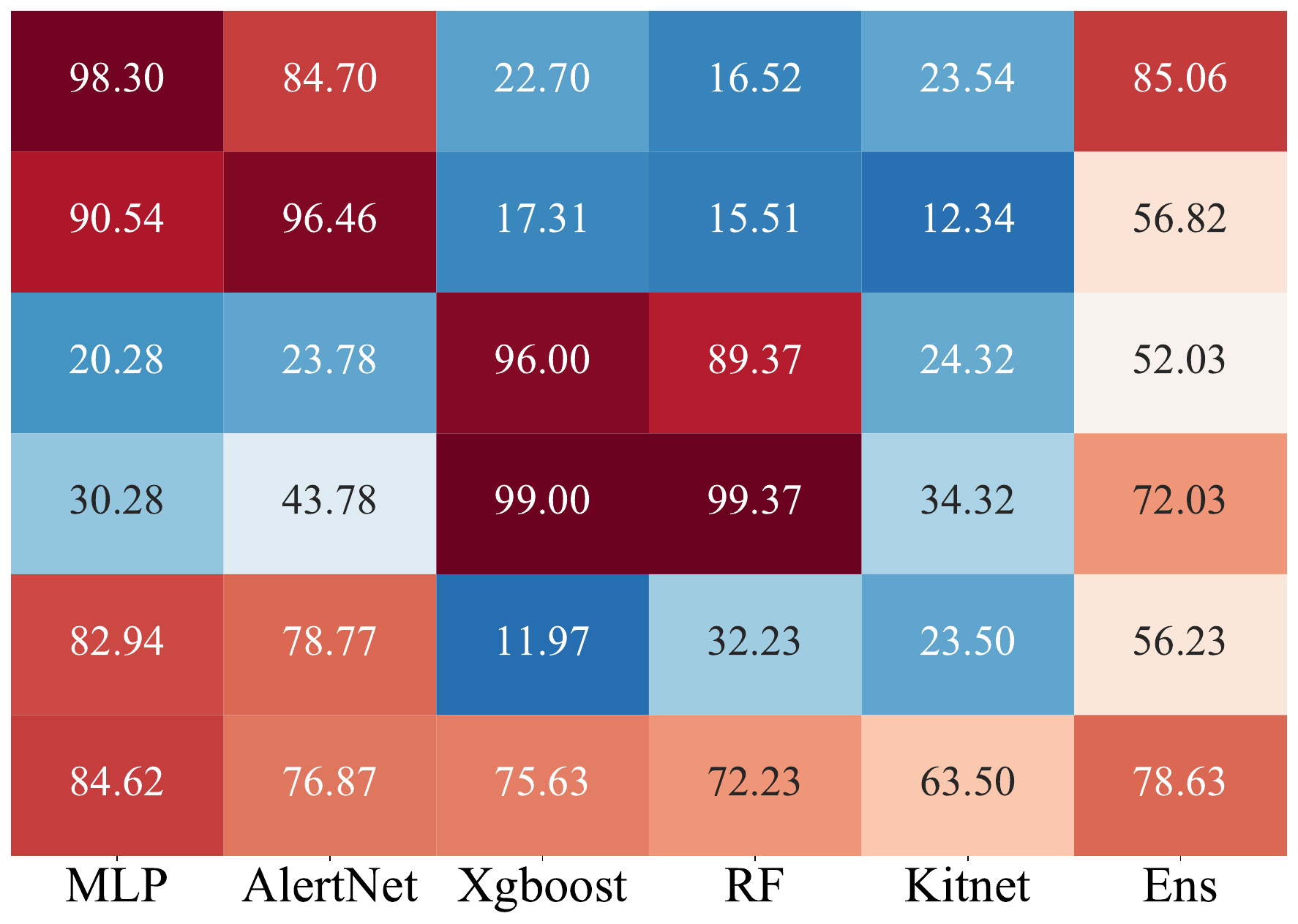}
  \caption{Kitsune-Mirai}
 \end{subfigure}
 \hfill
 \begin{subfigure}{.22\textwidth}
   \includegraphics[width=\textwidth]{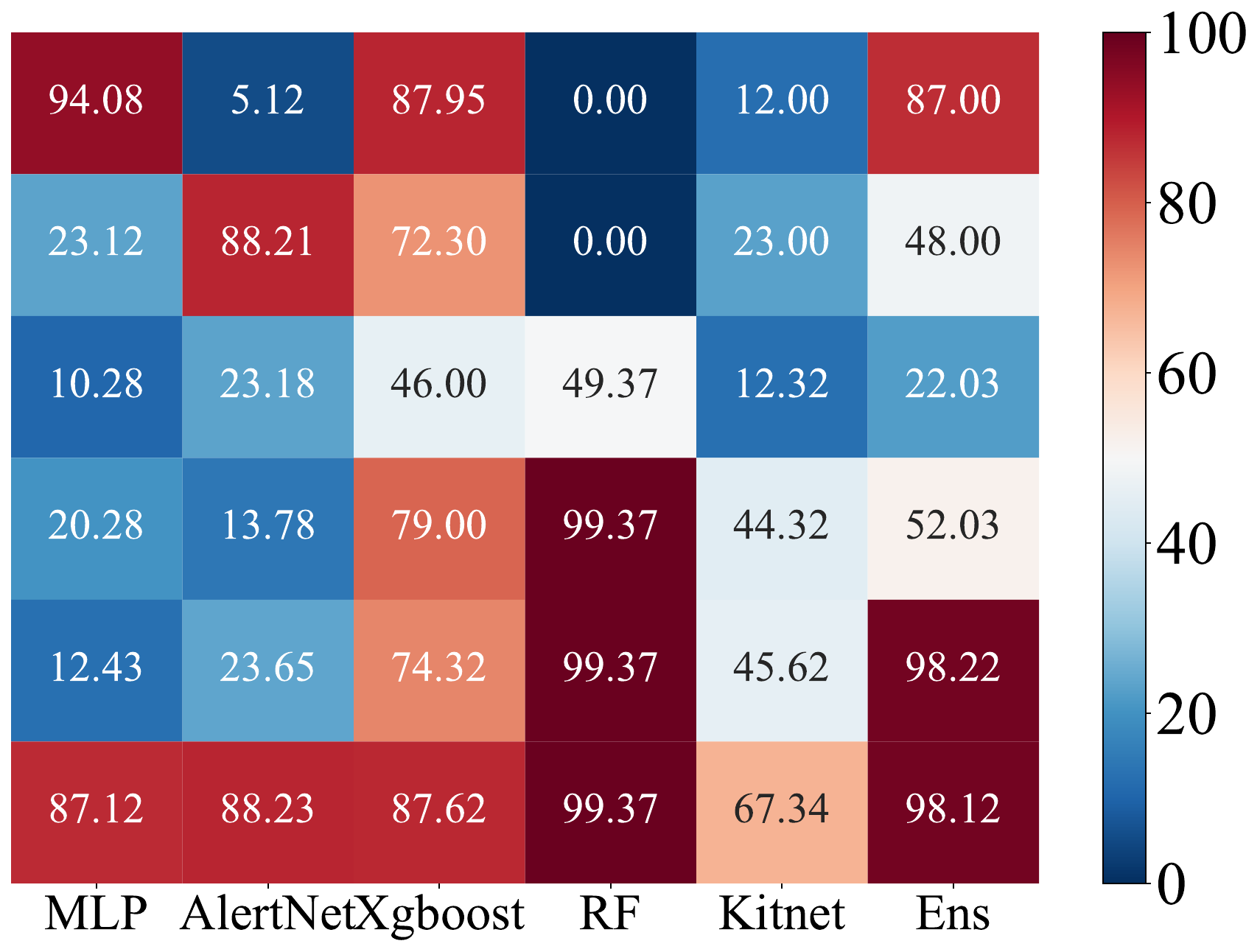}
   \caption{Kitsune-SYN-DoS}
 \end{subfigure}
 
 \caption{Cross-model transferability matrix: cell $(i, j)$ represents the success rate of attacks on classifier $j$ by the adversarial examples (AEs) generated for classifier $i$. Specifically, the rows indicate the substitute models that craft AEs, and the columns indicate the models under transfer-based attacks.}
 \label{fig-trans}

\end{figure*}
\begin{figure*}
  \centering
  \begin{subfigure}{.49\textwidth}
  \includegraphics[width=\textwidth]{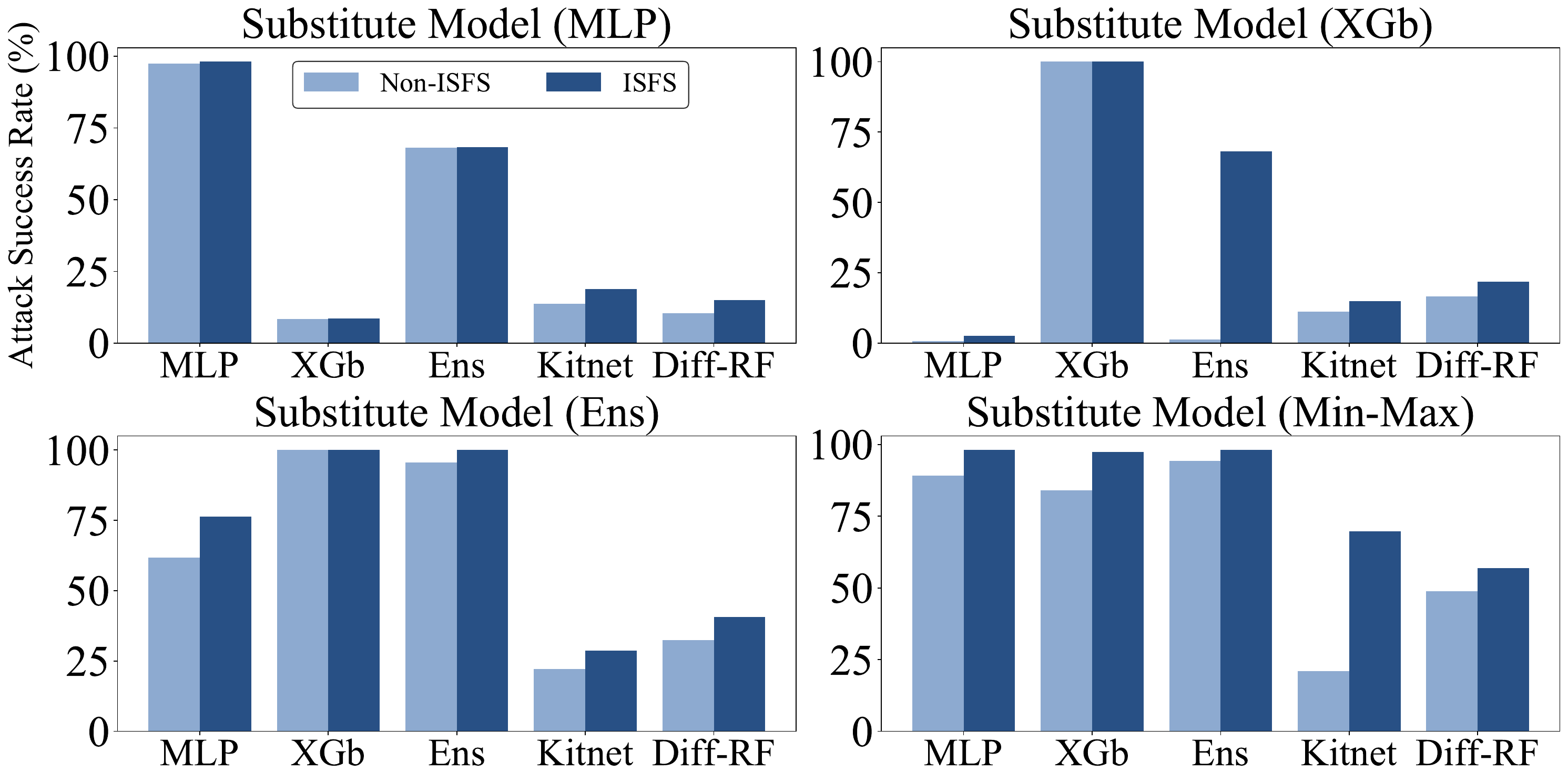}
  \caption{CicFlowMeter}
  \label{fig-feature-cic}
  \end{subfigure}
  \hfill
  \begin{subfigure}{.49\textwidth}
  \includegraphics[width=\textwidth]{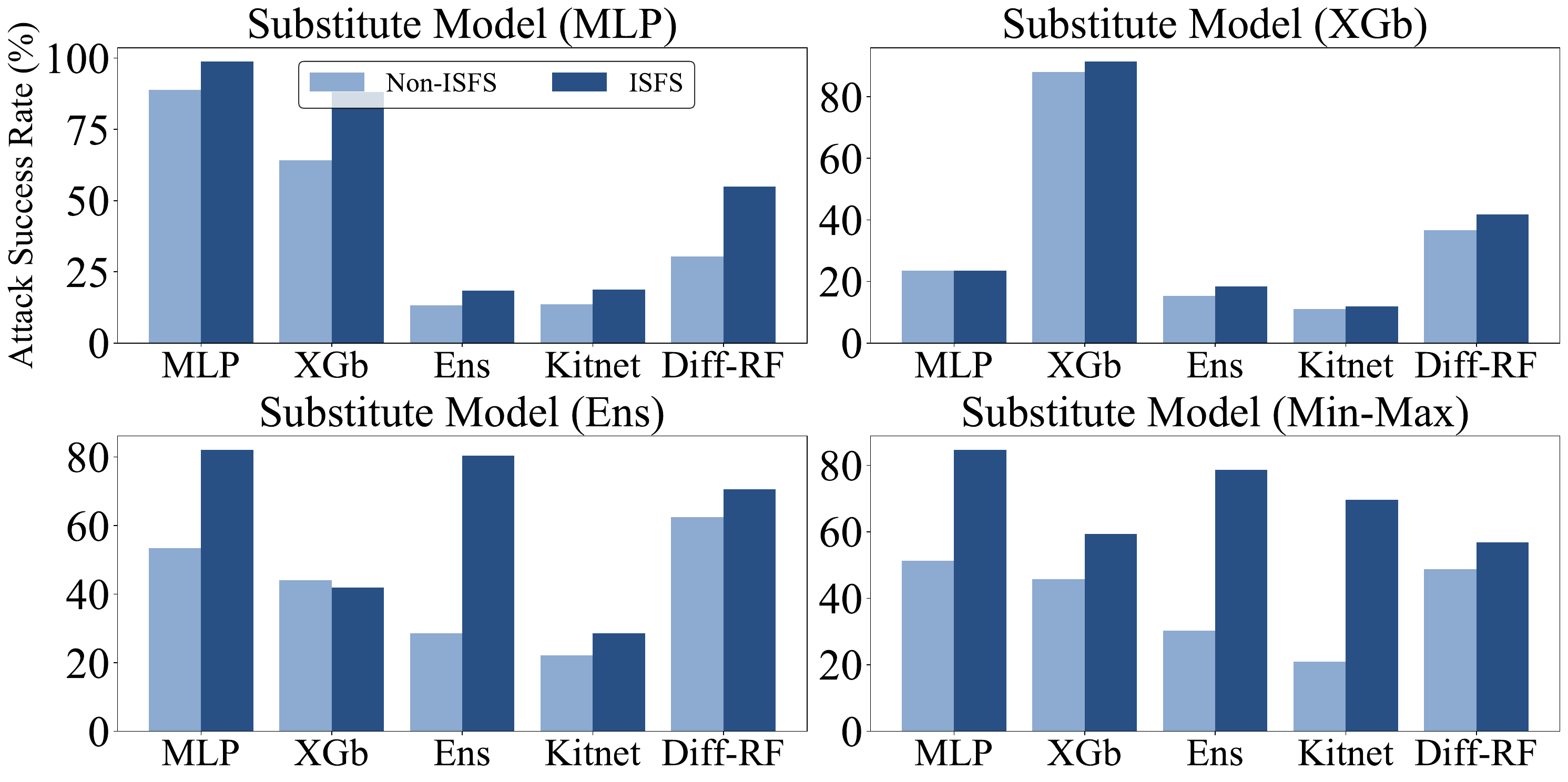}
  \caption{AfterImage}
  \label{fig-feature-after}
  \end{subfigure}
  \vspace{-0.05in}
  \caption{The impact of Important-Sensitive Feature Selection (ISFS) on the results of the ETA system.}
  \vspace{-0.15in}
  \label{fig-feature}
\end{figure*}

In this section, we compare the effects for the different components in our framework using two ablation studies.
\subsubsection{Impact of Substitute Model}
To explore the impact of different substitute models, we are interested in transferability across machine learning techniques.
Using the ETA attack method, we replace the substitute model with other models to investigate the impact of different substitute models on the results.
We build a cross-model transferability matrix where each cell $(i, j)$ represents the $ASR$ of using classifier $i$ to attack classifier $j$.
Specifically, the rows indicate the substitute models that craft AEs, and the columns indicate the models under transfer-based attacks.
As shown in Fig.~\ref{fig-trans},
We utilize MLP as a substitute model to attack other models.
\textcolor{black}{The results show that MLP can attack differentiable models like AlertNet more easily, achieving $ASR$ in 50\% in Average.}
The success rate of attacks against non-differentiable models (e.g., XGboost) is much lower, showing 0.64\% and 3.40\% $ASR$ in two IDS17 datasets.
Meanwhile, RF and Xgb are more powerful inter-transferability, but using them as substitute models to attack other models is less effective.
Generally, ensemble models have more excellent transferability.
Nevertheless, our Min-Max optimization model achieves 35\% to 100\% transferability and performs consistently, which shows that further optimization is necessary for the ensemble substitute model.

\subsubsection{Impact of ISFS}
To evaluate the effectiveness of our ISFS, we choose two datasets (IDS17-Botnet, Kitsune-Mirai) for our ablation studies.
Using $l_2$-norm, we limit the cost of the attack and stop the loop once the substitute model finds AEs. 
We want to test whether our ISFS can improve transferability.
\textcolor{black}{Fig.~\ref{fig-feature-cic} shows when the substitute is a single model (e.g., MLP), the use of ISFS has little impact on classical models (e.g., AlertNet) , but significant improvement for anomaly detection models (Kitnet, DEF-RF).}
When the substitute is an ensemble model (including Ens and Min-Max), there is an average improvement of 30\% in transferability using ISFS.
There is a similar effect for Kitsune-Mirai in Fig.~\ref{fig-feature-after}.
An additional benefit is that using ISFS means fewer features are modified, enabling the traffic-space constraints more easily.

\begin{figure*}
  \centering
  \begin{minipage}[b]{0.7\textwidth} 
  \begin{subfigure}{.32\textwidth}
      \includegraphics[width=\textwidth]{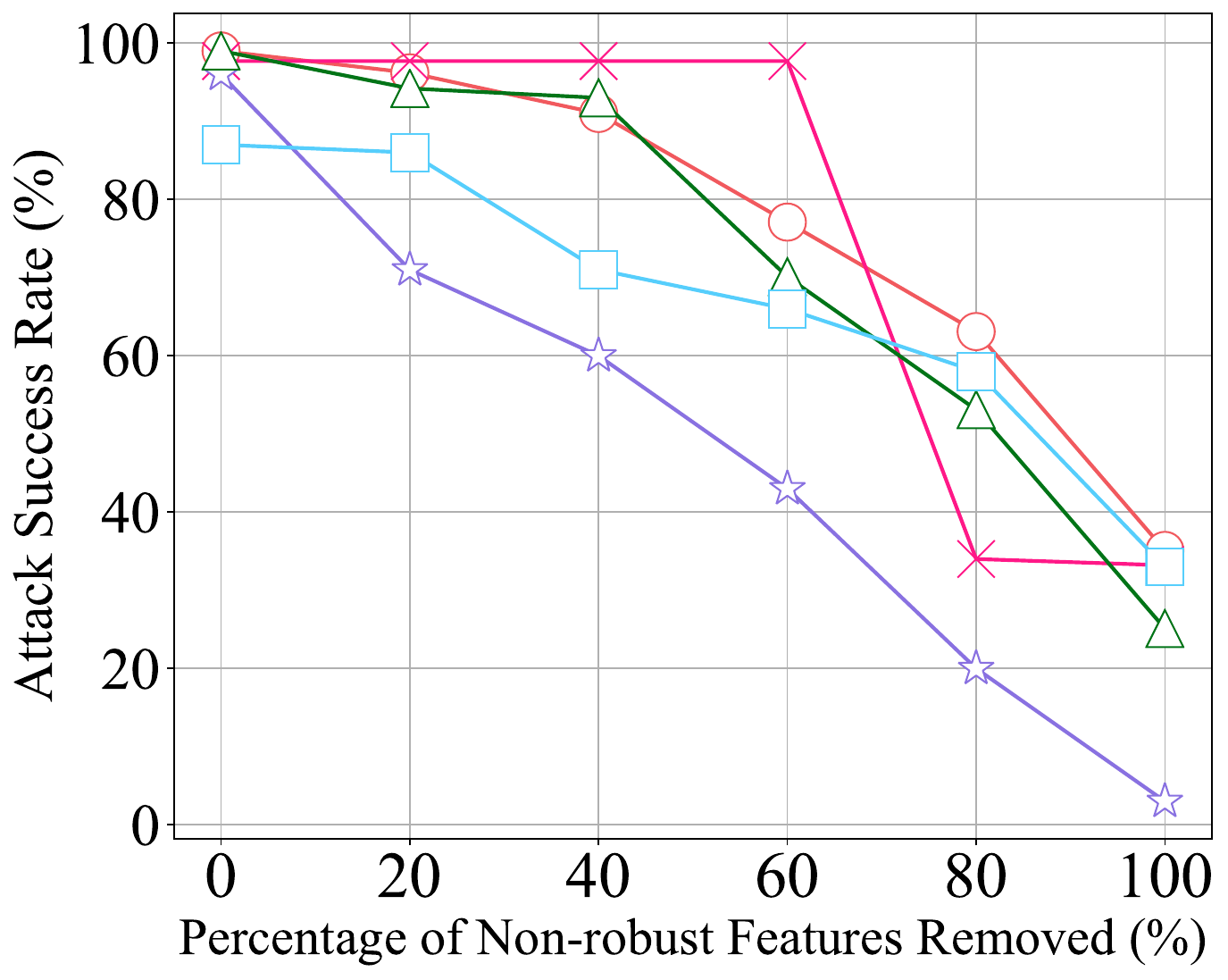}
      \caption{IDS17-Botnet-Remove}
      \label{fig-del-botnet}
  \end{subfigure}
  \hfill
  \begin{subfigure}{.32\textwidth}
      \includegraphics[width=\textwidth]{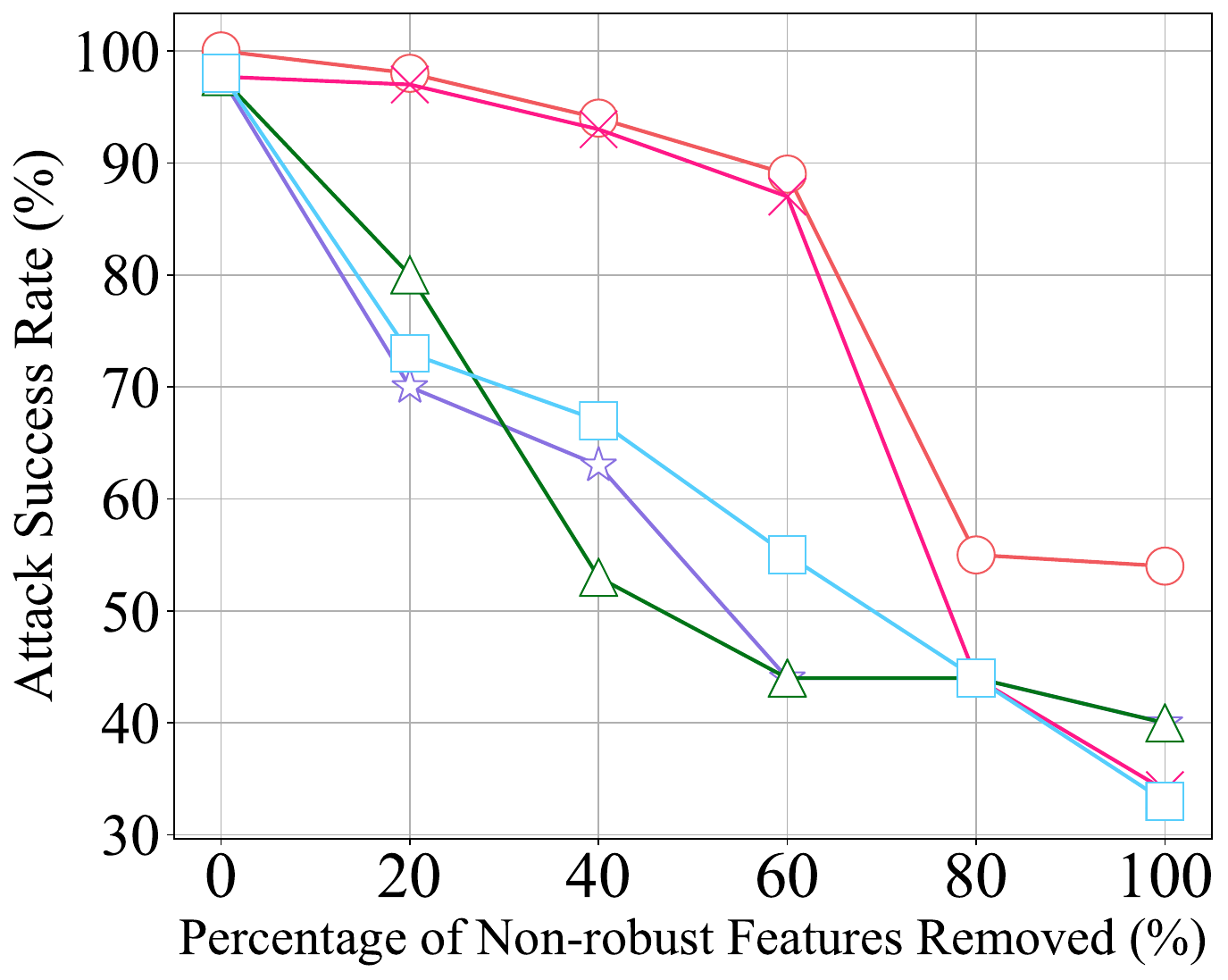}
      \caption{IDS17-DDoS-Remove}
      \label{fig-del-ddos}
  \end{subfigure}
  \hfill
  \begin{subfigure}{.32\textwidth}
      \includegraphics[width=\textwidth]{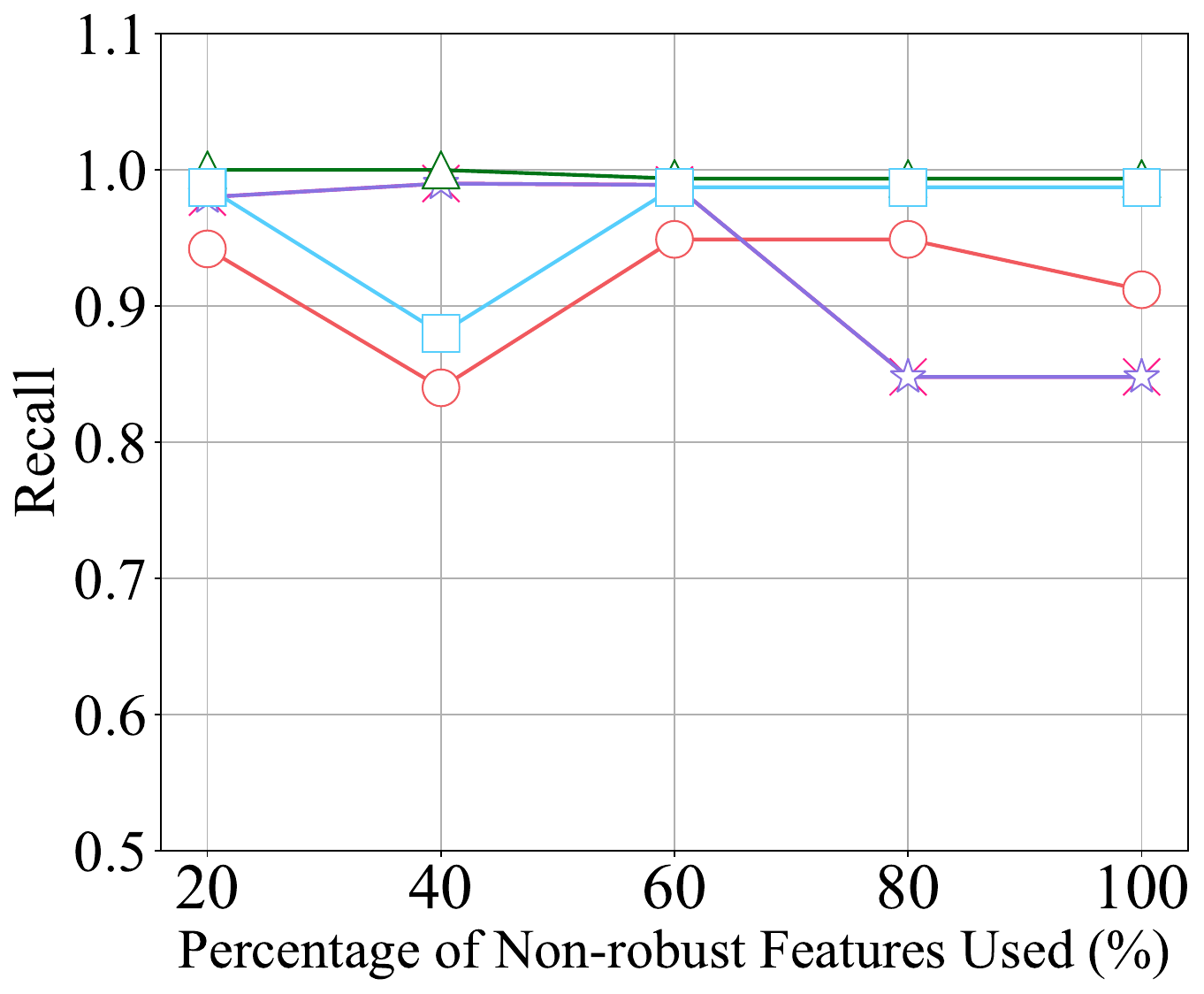}
      \caption{IDS17-Botnet-Use}
      \label{fig-use-botnet}
  \end{subfigure}

  \begin{subfigure}{.32\textwidth}
      \includegraphics[width=\textwidth]{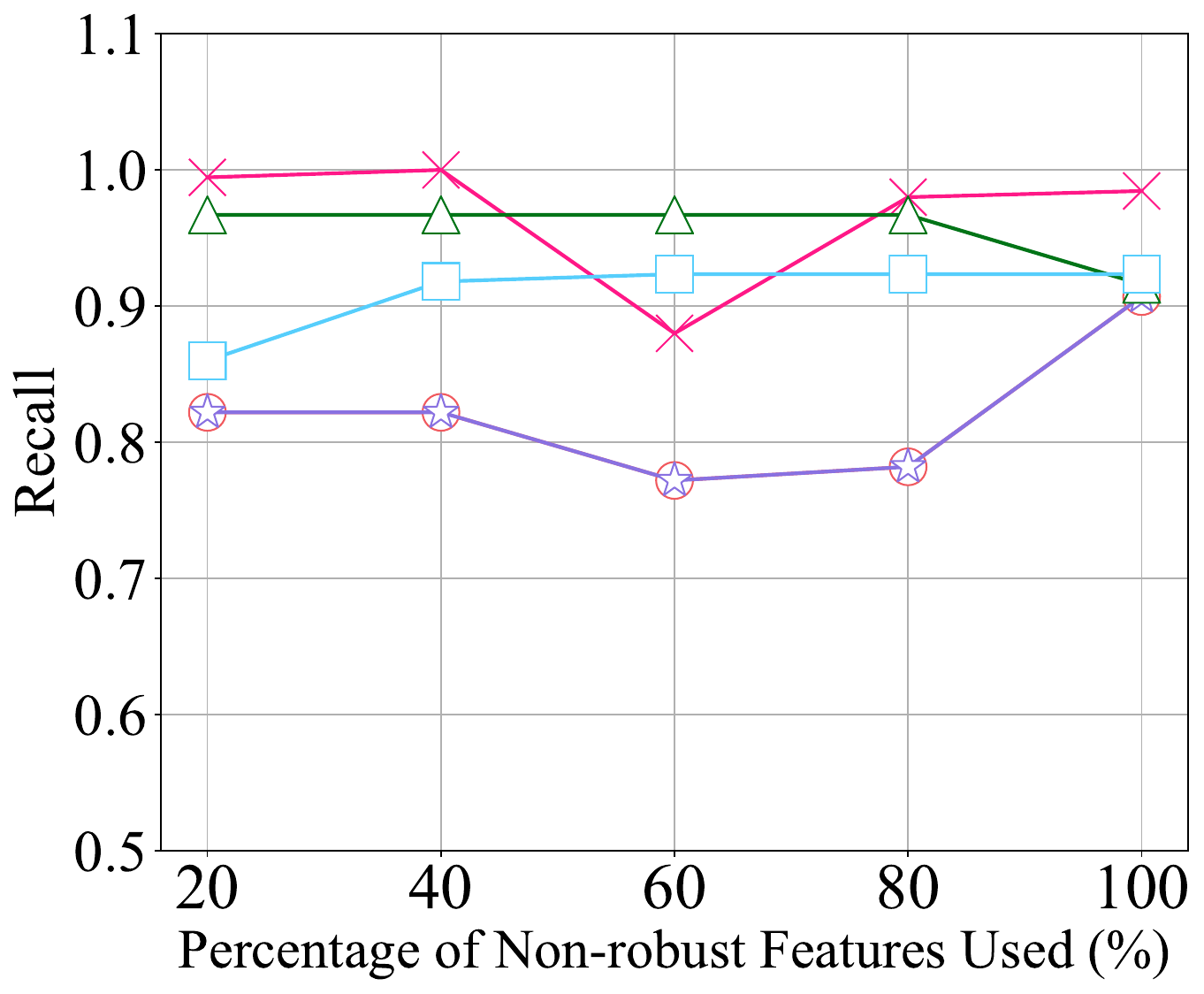}
      \caption{IDS17-DDoS-Use}
      \label{fig-use-ddos}
  \end{subfigure}
  \hfill
  \begin{subfigure}{.32\textwidth}
      \includegraphics[width=\textwidth]{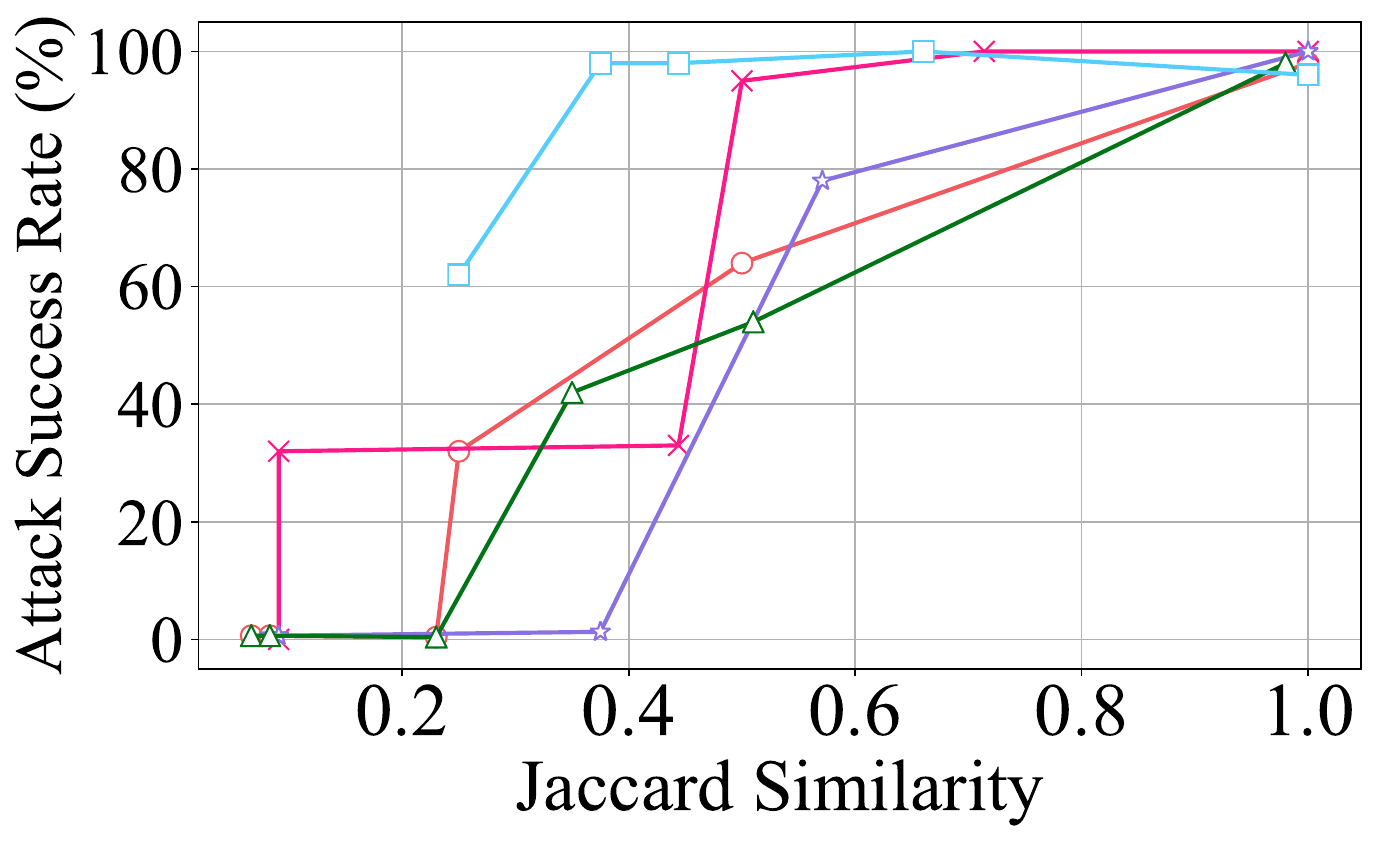}
      \caption{IDS17-Botnet-Compare}
      \label{fig-similar-botnet}
  \end{subfigure}
  \hfill
  \begin{subfigure}{.32\textwidth}
      \includegraphics[width=\textwidth]{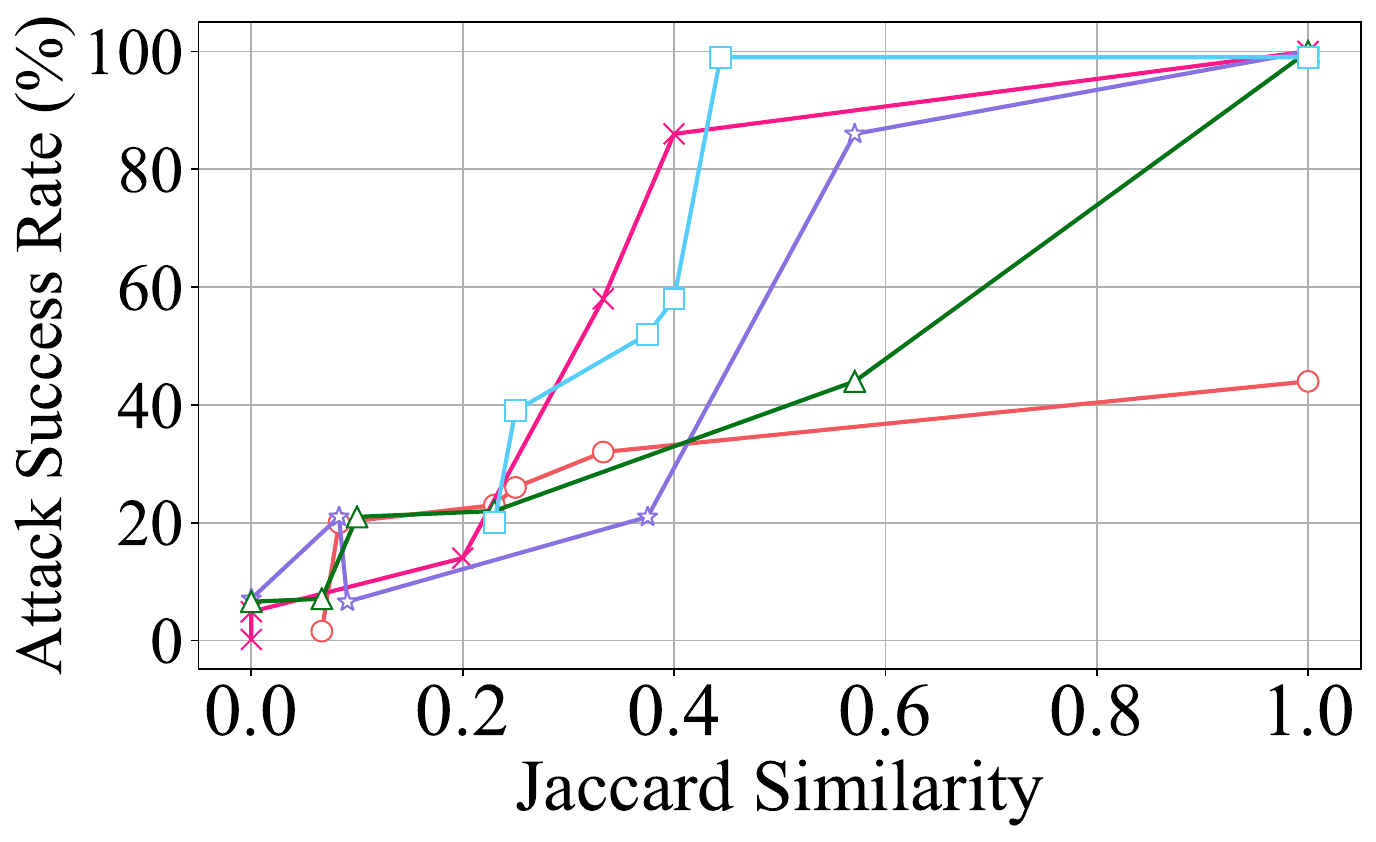}
      \caption{IDS17-DDoS-Compare}
      \label{fig-similar-ddos}
  \end{subfigure}
  \centering
  \begin{subfigure}{0.6\textwidth}
      \includegraphics[width=\textwidth]{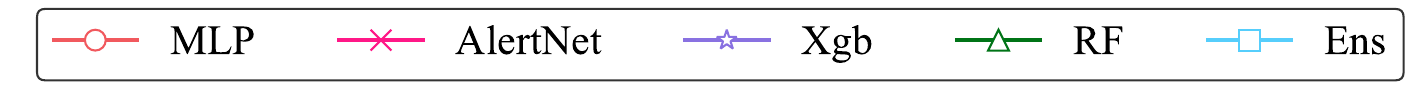}
  \end{subfigure}
 \caption{With removing, using, and comparing non-robust features, the experiment explains the reasons for the existence of adversarial examples (AEs) and adversarial transferability in ML-based NIDSs}
\end{minipage}
\hfill
\begin{minipage}[b]{0.25\textwidth} 
\includegraphics[width=\textwidth]{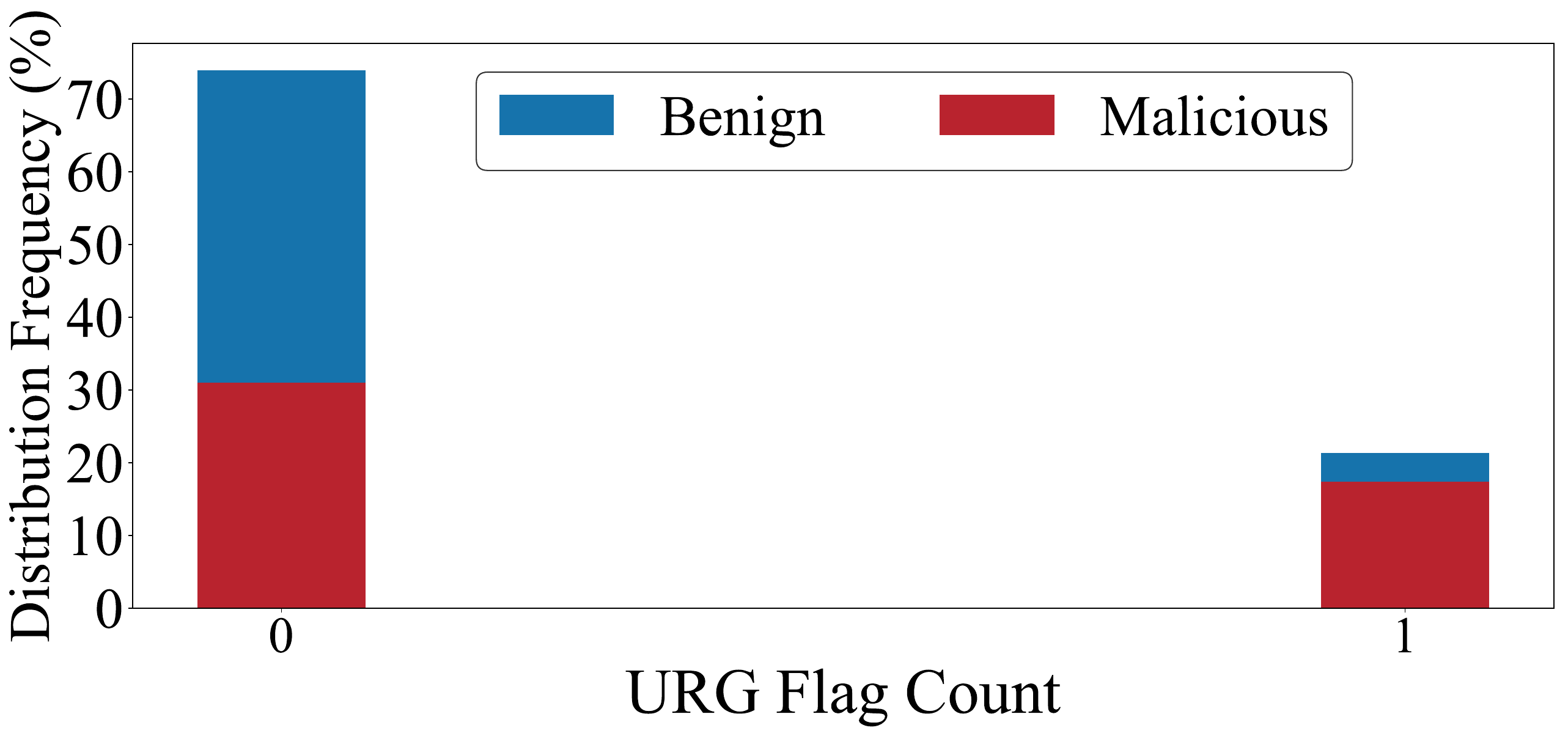}\\
\includegraphics[width=\textwidth]{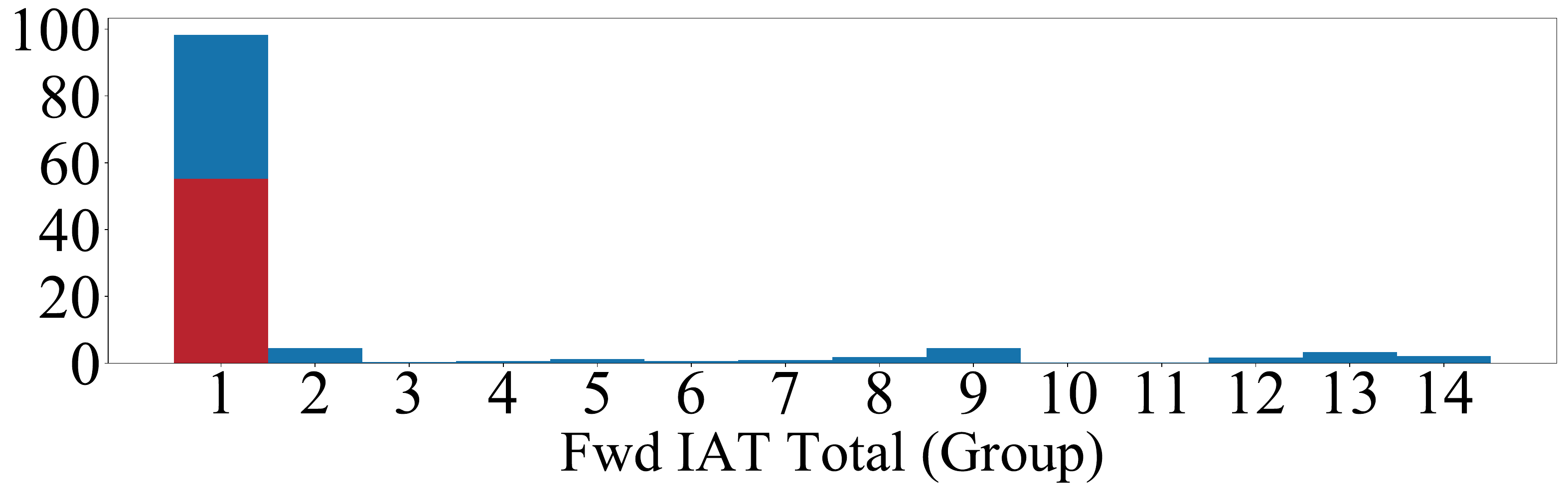}\\
\includegraphics[width=\textwidth]{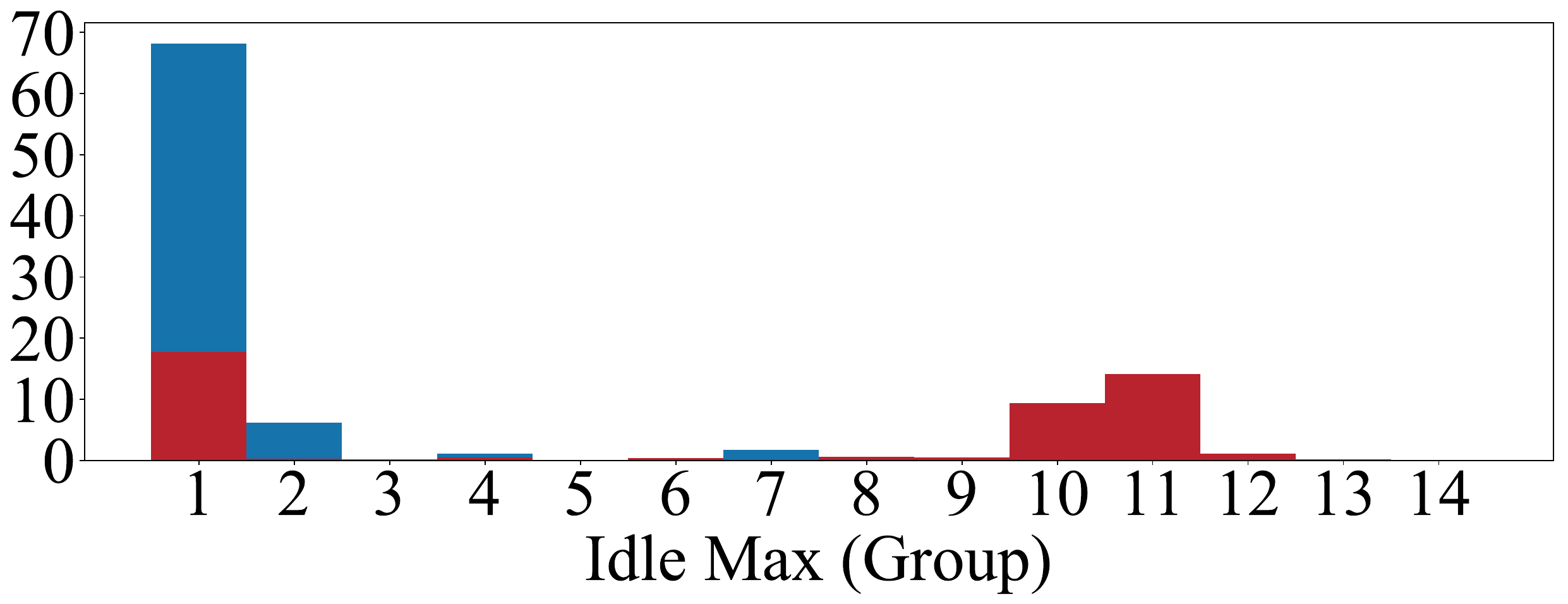}\\
\includegraphics[width=\textwidth]{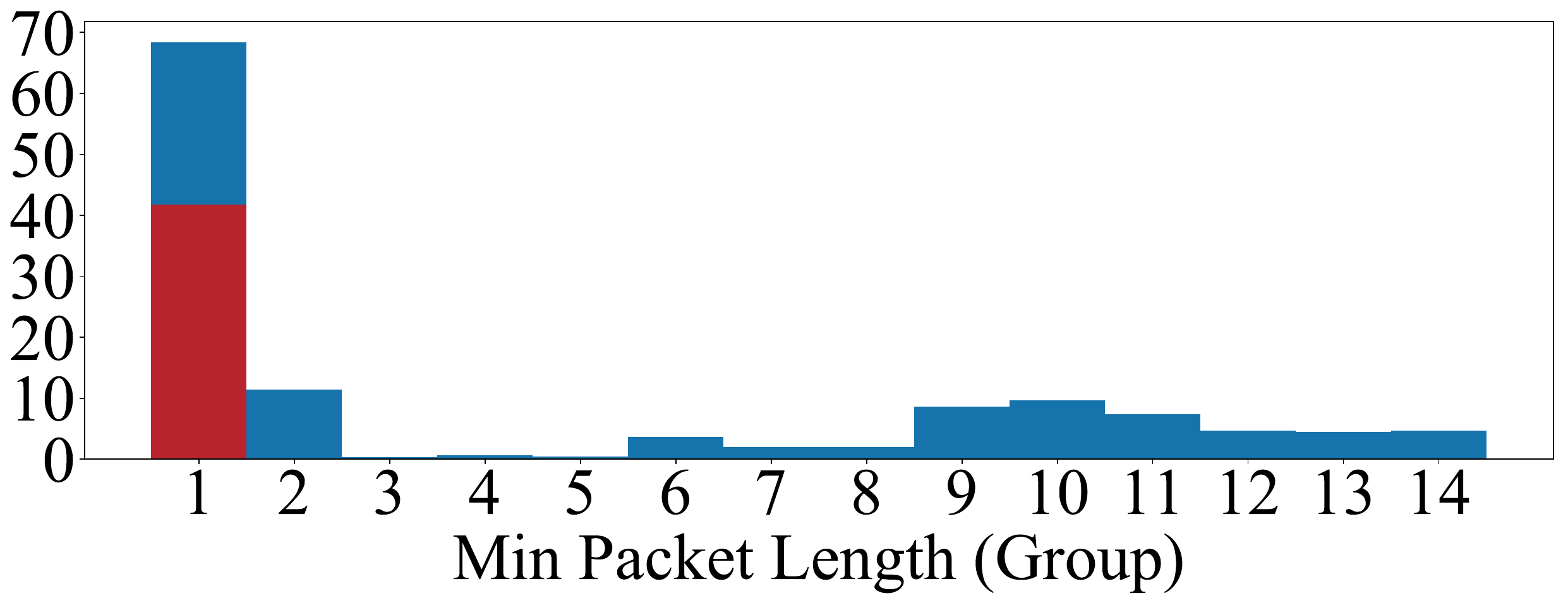}
\caption{Frequency Distribution Histogram}
\label{fig-fdh}
\end{minipage}

\end{figure*}
\subsection{Interpretation}
\label{subsec-interpretation}
As discussed in Section~\ref{subsec-interpretation-eta}, we claim that adversarial vulnerability is a direct result of the existence of features that are important and sensitive (\emph{non-robust features}).
The adversarial transferability is due to similar non-robust features learned by the two models.
In this section, we conduct three experiments and two cases to verify our interpretation.
Meanwhile, two significant misconceptions with regards to the current use of machine learning in NIDSs systems are uncovered in~\ref{subsec-use-non-robust} and~\ref{subsec-understand-case}.

\subsubsection{Impact of Deleting Non-robust Features}
To illustrate the effect of non-robust features on the results, we continuously remove non-robust features and then conduct an adversarial attack on the model to assess the robustness of the model.
As non-robust features are removed from the model, the attack success rate ($ASR$) gradually decreases, indicating the model is more robust in Fig.~\ref{fig-del-botnet} and Fig.~\ref{fig-del-ddos}.
The x-axis is increasing order of deleted features.
While the attack barely succeeds by removing all non-robust features, it proves that only containing robust features leads to significantly more robust classifiers.
Consequently, the experimental results can generally support the assertion that non-robust features are responsible for AEs.
\begin{table*}[]
    \caption{EXAMPLE INTERPRETATIONS FOR CICIDS2017} 
    \vspace{-0.1in}
    \label{table-case}
    \begin{center}
  
    \begin{subtable}{.9\textwidth}
        \caption{Ground truth: Botnet}
        \begin{tabular}{@{}lllll@{}}
            \toprule
            Feature Name       & Feature Meaning                                                                               & robust & FAI  & FS   \\ \midrule
            URG Flag Count     & The number of flag for emergency data                                                         & Yes    & 0.1000 & 0.3514 \\
            PSH Flag Count     & The number of PSH flag & Yes    & 0.1750 & 0.6671 \\
            Packet Length Mean & The mean of packet length                                                                       & Yes    & 0.0250 & 0.0500 \\
            Flow Duration      & The duration of the flow                                                                      & No     & 0.0250 & 1.0000 \\
            Fwd IAT Total      & Flow Inter Arrival Time, the time between two packets sent in either direction (total)        & No     & 0.0125 & 1.0000 \\
            idle MIn           & The amount time of a flow is idle before it becomes active (min)                          & No     & 0.0125 & 1.0000 \\ \bottomrule
            \end{tabular}
    \end{subtable}
    \begin{subtable}{.9\textwidth}
        \caption{Ground truth: DDoS}
        \begin{tabular}{@{}lllll@{}}
            \toprule
            Feature Name          & Feature Meaning                                                                               & robust & FAI  & FS   \\ \midrule
            URG Flag Count        & The number of flag for emergency data                                                         & Yes    & 0.0500 & 0.6000 \\
            ACK Flag Count        & The number of ACK flag & Yes    & 0.0250 & 0.3600 \\
            Idle Max              & The amount time of a flow is idle before it becomes active (max)                                                                       & Yes    & 0.0125 & 0.1800 \\
            Min Packet Length     & The min of packet length                                                                    & No     & 0.1250 & 0.9950 \\
            Fwd IAT Total         & Flow Inter Arrival Time, the time between two packets sent in either direction (total)        & No     & 0.0250 & 0.9950 \\
            Down/Up Ratio         & The ratio of up to down packets                         & No     & 0.0050 & 0.9900 \\ \bottomrule
            \end{tabular}
    \end{subtable}
    \end{center}
    \end{table*}
\begin{figure*}
    \centering
    \begin{subfigure}{.33\textwidth}
        \includegraphics[width=\textwidth]{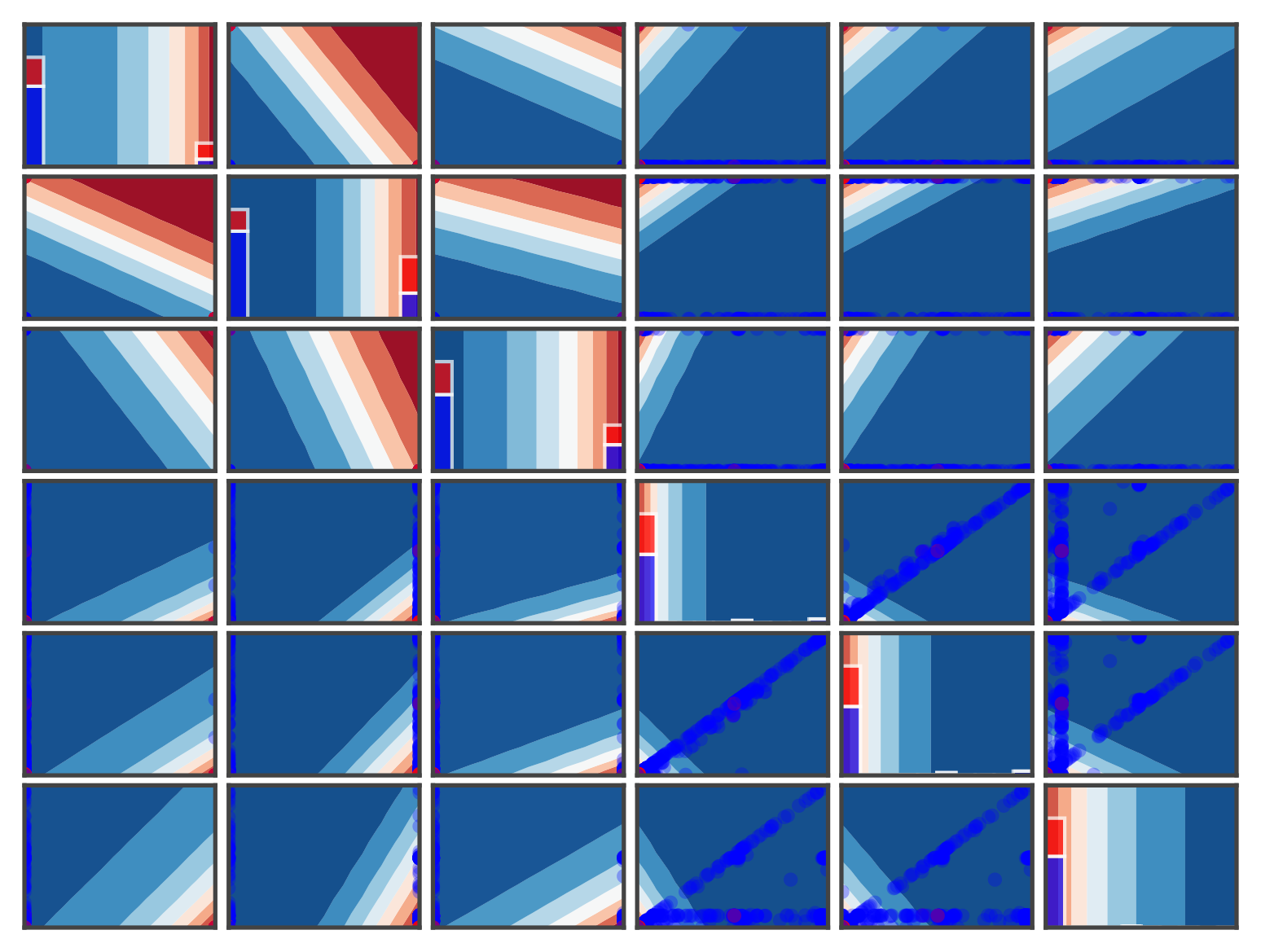}
        \caption{Botnet-MLP}
        \label{fig-dim-botnet-mlp}
    \end{subfigure}
    \hfill
    \begin{subfigure}{.33\textwidth}
        \includegraphics[width=\textwidth]{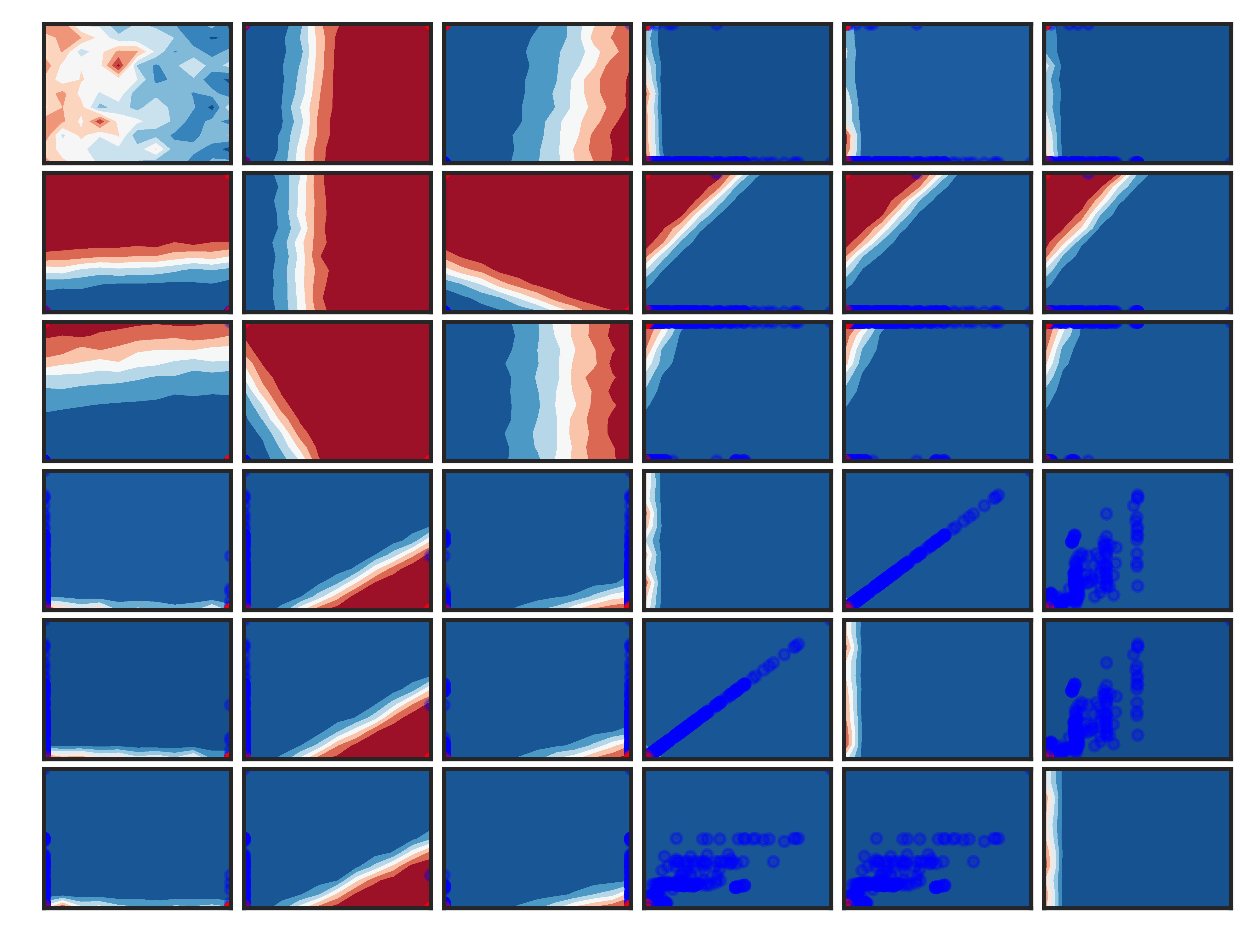}
        \caption{Botnet-AlertNet}
        \label{fig-dim-botnet-alertnet}
    \end{subfigure}
    \hfill
    \begin{subfigure}{.33\textwidth}
        \includegraphics[width=\textwidth]{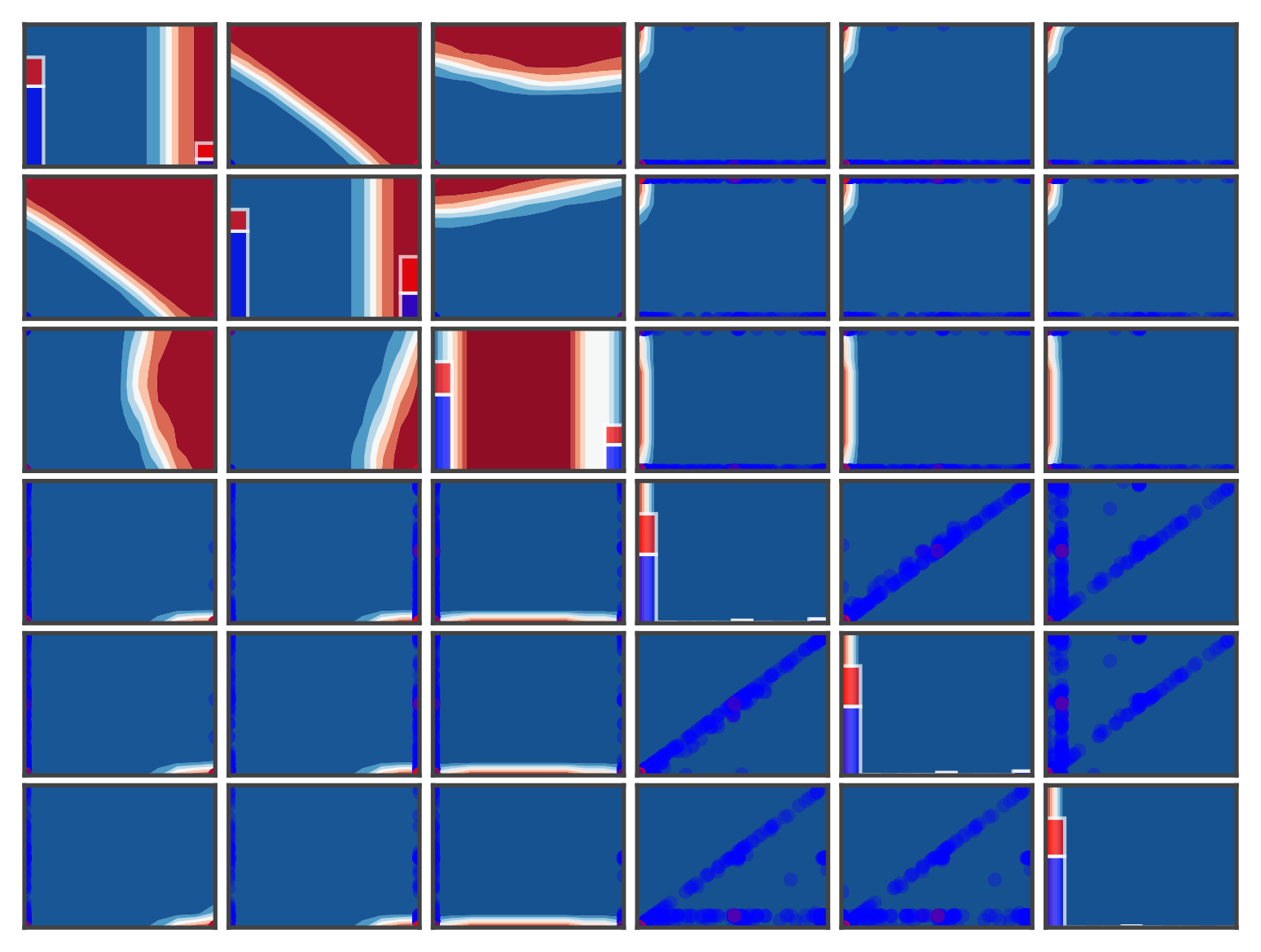}
        \caption{Botnet-IDSNet}
        \label{fig-dim-botnet-idsnet}
    \end{subfigure}
    \hfill
    \begin{subfigure}{.33\textwidth}
    \includegraphics[width=\textwidth]{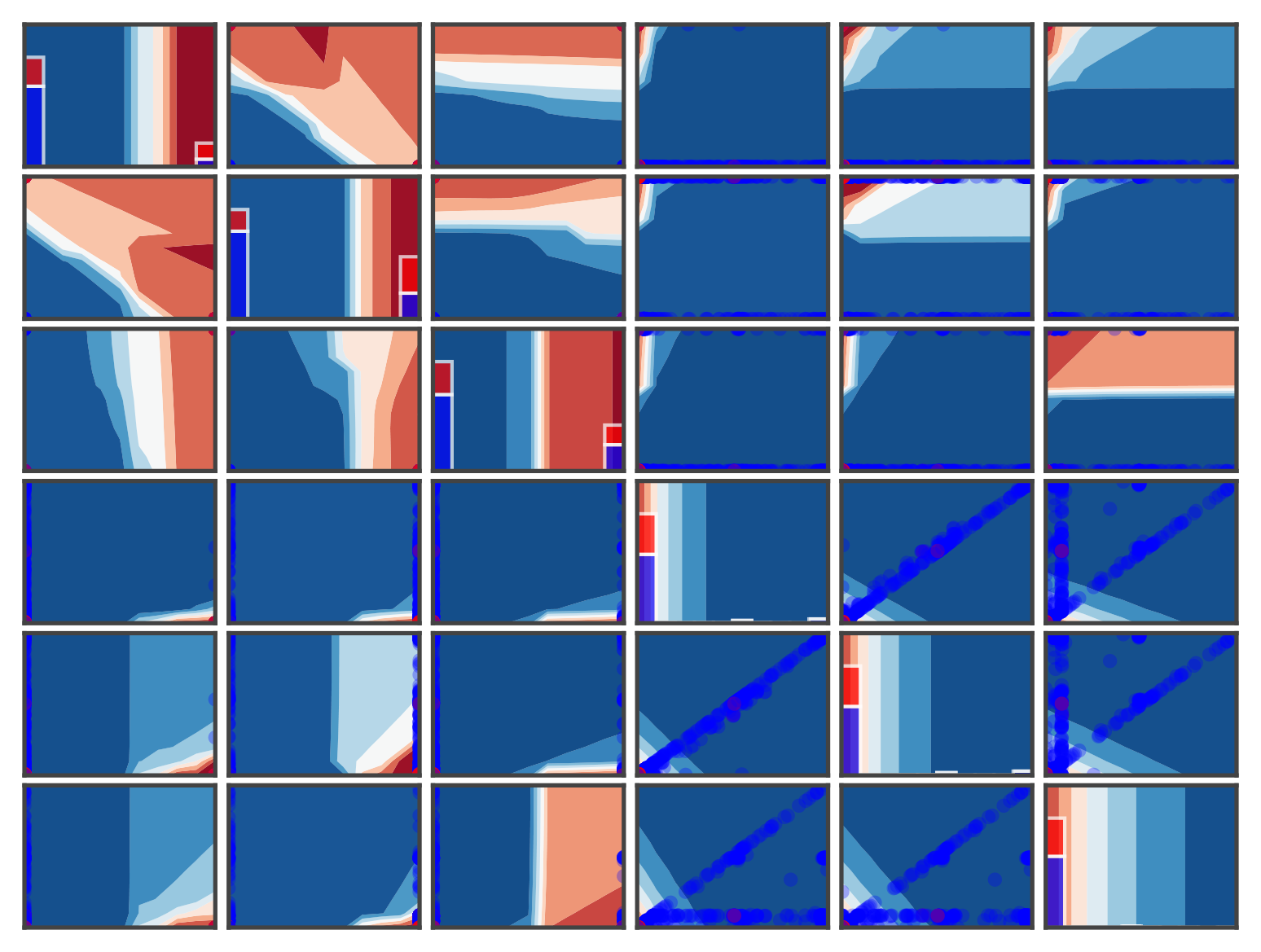}
    \caption{Botnet-Ens}
    \label{fig-dim-botnet-ens}
    \end{subfigure}
    \hfill
    \begin{subfigure}{.33\textwidth}
    \includegraphics[width=\textwidth]{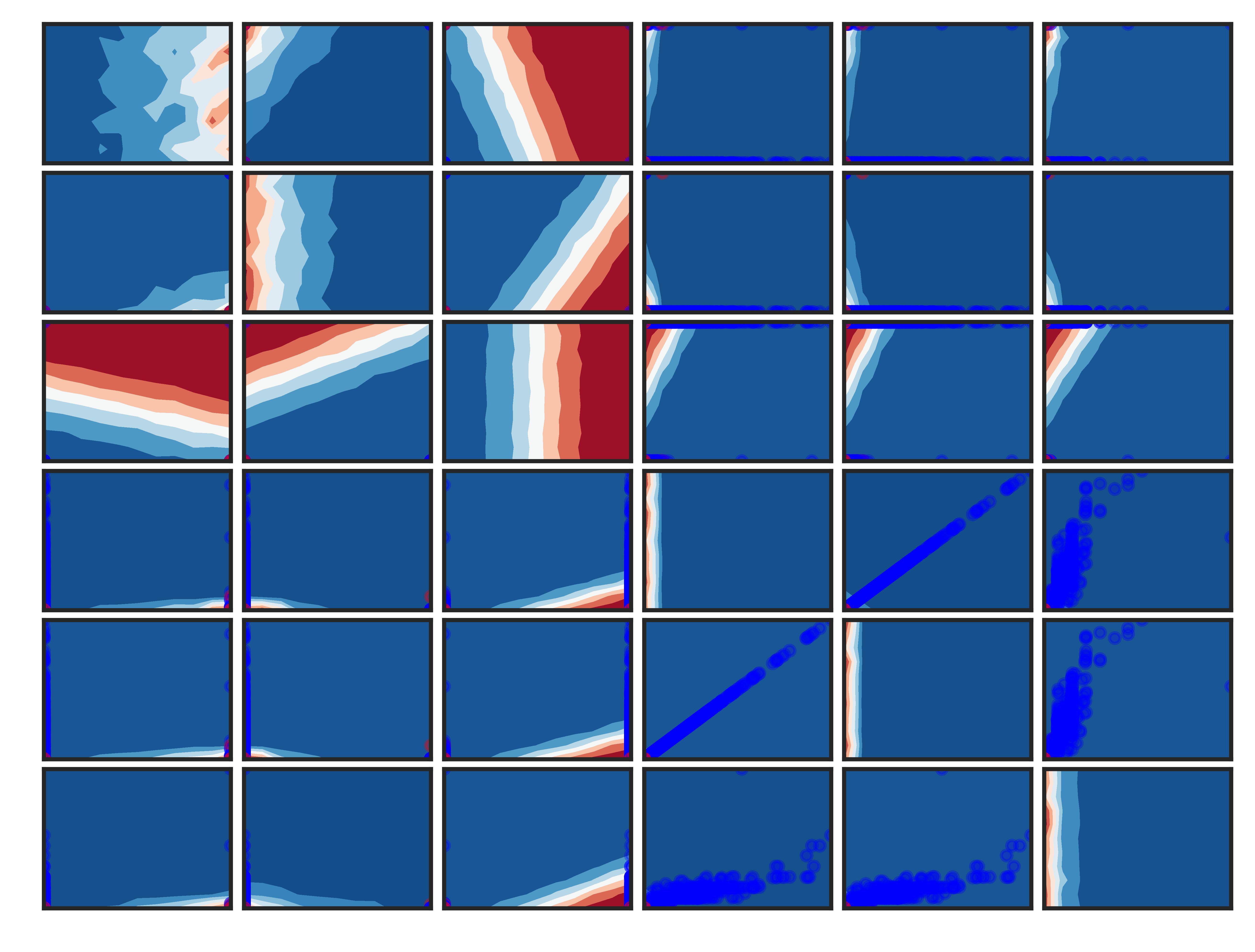}
    \caption{DDos-AlertNet}
    \label{fig-dim-ddos-alertnet}
    \end{subfigure}
     \hfill
    \begin{subfigure}{.33\textwidth}
    \includegraphics[width=\textwidth]{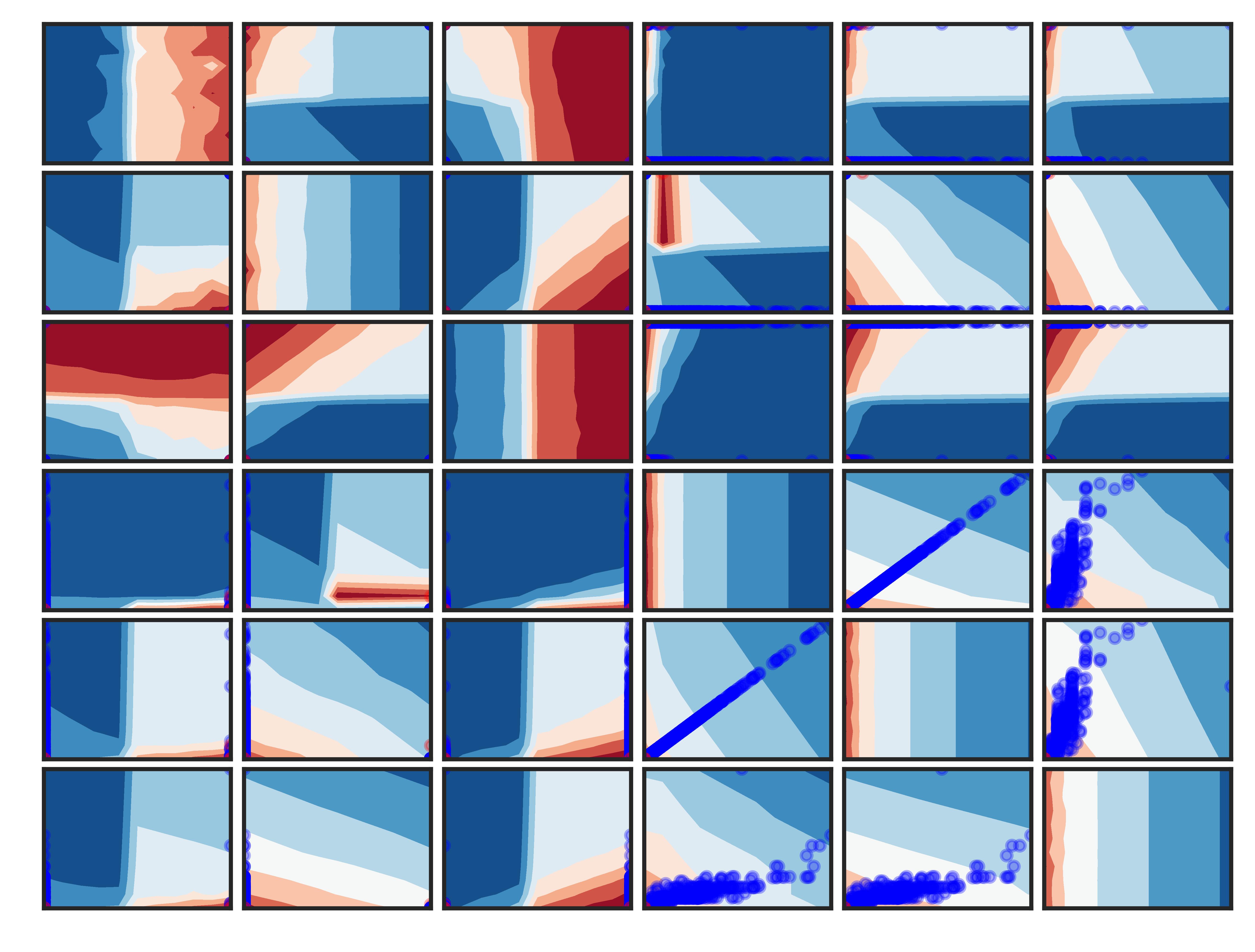}
    \caption{DDos-Ens}
    \label{fig-dim-ddos-ens}
    \end{subfigure}
    \caption{High-dimensional decision boundaries of different models: the first three features are robust, and the last three are non-robust; The red area indicates that the classifier judged the samples as malicious, blue as benign; Each subplot represents a projected decision boundary between two features on models.}
    \label{fig-dim}
    \end{figure*}
\subsubsection{Impact of Only Using Non-robust Features}
\label{subsec-use-non-robust}
Additionally, we conducted a corresponding experiment using only non-robust features to train the model.
The result is shown in Fig.~\ref{fig-use-botnet} and~\ref{fig-use-ddos}.
A surprising result shows that if we train the model using only non-robust features, the model also achieves high accuracy and is insensitive to the dimensionality of the non-robust features.
This suggests that non-robust features take on a significant role in the learned classifier. 
On the other hand, our experiments also reveal a major misconception about applying machine learning to NIDSs systems--\emph{focusing solely on model accuracy is inappropriate}.
The specific goal of training classifiers is to maximize their accuracy.
Consequently, classifiers tend to use any available features to distinguish samples, even meaningless features.
Models with high accuracy have also likely learned these non-robust features.
Therefore, the general approach to training machine learning, which focuses only on accuracy, is inappropriate.
Instead, the correct training method should take into account both model accuracy and robustness.

\subsubsection{The Relationship between Similarity of Non-robust Features and Adversarial Transferability}
An intriguing phenomenon in the AEs is that the perturbations computed for one model are often transferred to other independently trained models.
We demonstrate that this phenomenon can be directly attributed to the natural consequence of similar non-robust features.
Below, we compare the relationship between the similarity of the non-robust features and adversarial transferability.
We cast five base models (MLP, ALertNet, Xgb., RF, Ens.) as substitute models to verify this correlation and leverage Jaccard Similarity (JS) to measure the similarity of two non-robust features sets $S_1,S_2$, which is defined as $|S_1\cap S_2| /|S_1 \cup S_2|$.
In Fig.~\ref{fig-similar-botnet}, the x-axis indicates the Jaccard Similarity (JS) of non-robust feature sets between substitute and target model.
The y-axis represents the $ASR$ of using substitute models to attack target models.
Using the AlertNet model as an example, we calculate the JS values between the other and AlertNet models.
As shown in Fig.~\ref{fig-similar-botnet}, when the JS value is 0.42, the $ASR$ from AlertNet to target model is 35\%, and when the JS value is 0.5, the $ASR$ is 92\%.
Generally, Fig.~\ref{fig-similar-botnet} and~\ref{fig-similar-ddos} show that adversarial transferability and similarity of non-robust features exhibit a positive correlation.
Thus, adversarial transferability results from similar non-robust features in the two models.

\subsubsection{Understanding the Impact of Non-robust and robust Features}
\label{subsec-understand-case}

We use two cases (Botnet and DDoS) to understand how non-robust and robust features affect model decisions.
Three robust and non-robust features are selected to train the model, and their names and meanings are listed in Table~\ref{table-case}.
Meanwhile, we plot their decision boundaries as displayed in Fig.~\ref{fig-dim-botnet-mlp} and~\ref{fig-dim-botnet-alertnet}.
The first three features are robust, and the last three are non-robust.
Each subplot represents a projected decision boundary between two features on the different models.
Obviously, there is a clear difference between the decision boundaries of robust and non-robust features.
Specifically, we find that the robust features contain more feature interactions and the classification regions are more evenly distributed, while the classification regions learned by the non-robust feature is essentially the benign region.

Next, we will examine why robust and non-robust features have such different decision boundaries.
We choose a robust and non-robust feature to see how they impact the results.
For example, UFC (URG Flag Count) is a robust feature, which rarely occurs in normal settings, indicating a flag of emergency data presence.
We plot the FDH (Frequency Distribution Histogram) of UFC, which contains only two values, 0 and 1 in the x-axis.
The y-axis indicates the distribution frequency of samples in the group.
As can be seen from Fig.~\ref{fig-fdh}, a majority of samples with a UFC value of 1 are malicious.
Furthermore, changing the UFC value to 0 will not allow the sample to be classified as benign because both malicious and benign samples have a high probability of occurrence when UFC is 0.
The two properties indicate that robust features we find can describe malicious traffic patterns.
In contrary, FIT (Fwd IAT Total) is a non-robust feature, which indicates the time between the arrival of two forwarding packets.
We divide feature values into 14 groups equally, with the x-axis indicating group number.
The FIT values for malicious samples are only distributed in one group, while they are distributed across all groups for benign samples.
Therefore, we only need to make a small perturbation in FIT to classify the malicious samples as benign.
Plotting the FDH of another robust feature IM (Idle Max) and MPL (Min Packet Length) in Fig.~\ref{fig-fdh}, we generally understand that the large differences in the distribution of these two types of features lead to differences in decision boundaries.

\textcolor{black}{We continue to examine whether the properties of this non-robust decision boundary across the model still exist.
Specifically, we train the IDSNet and Ens models using the same six-dimensional features as MLP model.
As shown in Fig.~\ref{fig-dim-botnet-idsnet} and~\ref{fig-dim-botnet-ens}, we find that most IdsNet and Ens classification regions are benign using these non-robust features to train models.}
It means these non-robust features are a vulnerability in feature extractors.
In contrary, with robust feature training models, every model learns decision boundaries with reliable decision capabilities.
This finding demonstrates the second misconception about applying machine learning to NIDSs system--\emph{choosing intrinsic features is more important than optimizing the model in current ML-based NIDSs}.  
Using these non-robust features to train the model, we will not learn the intrinsic traffic patterns no matter how to optimize the model.
In addition, these non-robust features make the model vulnerable to being evaded by attackers.
Therefore, for current ML-based NIDSs, it is more important to choose the essential features rather than to optimize the model indefinitely.

\subsection{Hyper-parameters Sensitivity}
\label{subsec-hyper}
ETA has some hyper-parameters in Gradient Evaluation and ISFS.
Below, we test the sensitivity of these hyper-parameters.
\begin{table}[]
    \caption{HYPER-PARAMETERSENSITIVITY} 
    \begin{subtable}{.45\textwidth}
        \caption{Hyper-parameters in Gradient Evaluation}
            \label{table-hyper-GE}
        \begin{tabular}{@{}lllll@{}}
            \toprule
            Parameters & default & Testing Range & ASR(range) & ASR(std.) \\ \midrule
            $\sigma$       & 20        &[10,50]   & [92.01\%-93.46\%]   &0.0053           \\
            $q$       &0.05         &[0.01,0.09]   & [92.01\%-93.86\%]  &0.0088           \\ 
            $\zeta$ &100 &[25,150] & [93.42\%,93.42\%] &0 \\
            $max_{iter}$ &20 &[5,30] &[91.03\%-97.98\%] &0.0285 \\
            \bottomrule
            \end{tabular}
    \end{subtable}

    \begin{subtable}{.45\textwidth}
        \caption{Hyper-parameters in ISFS}
        \label{table-hyper-ISFS}
        \begin{tabular}{@{}lllll@{}}
            \toprule
            Parameters & default & Testing Range & ASR(range) & ASR(std.) \\ \midrule
            $\tau$       & 10        &[4,12]  & [89.92\%,92.02\%]            &0.0148           \\
            $\epsilon$       &0.33         &[0.01,0.66]   & [89.5\%,90.02\%]      &0.0095           \\ 
            \bottomrule
            \end{tabular}
    \end{subtable}

\end{table}

\noindent\textbf{Hyper-parameters in Gradient Evaluation.}
We evaluate four hyper-parameters in Gradient Evaluation.
They are search variance $\sigma$, number of samples $q$, variance of gaussian distribution $\zeta$ and, number of maximum iteration $max_{iter}$.
Testing is done by varying one parameter with a reasonable range while fixing other parameters every time. 
Table~\ref{table-hyper-GE} lists their value range for testing, as well as default values when testing other parameters.
The results of the change and standard deviation of $ASR$ are also listed in Table~\ref{table-hyper-GE}.
We can find the change of ASR is small (e.g., std. for all parameters is $<$1\%). 
Thus, the performance of Gradient Evaluation is basically insensitive to hyper-parameters.

\noindent\textbf{Hyper-parameters in ISFS.}
We evaluate two hyper-parameters in ISFS.
They are the number of important features $\tau$ and sensitive threshold $\epsilon$.
From the results in Table~\ref{table-hyper-ISFS}, when the number of important features changes from 4 to 12, the ASR changes within 2\%.
We also test the effect of the sensitivity threshold $\epsilon$ on the ASR, and the experimental results find that $\epsilon$ also has little effect on the ASR.
These results indicate that the ISFS algorithm is not sensitive to hyper-parameter $\tau$ and $\epsilon$. 

\subsection{Evaluations in Real Environment}
\label{subsec-real}
\textcolor{black}{
We have deployed an intrusion detection system powered by machine learning algorithms on the enterprise network testbed in our lab. The network consists of 5 Windows and 7 Linux servers configured to simulate normal user activities as well as malicious activities. To facilitate robustness testing, we leveraged Tshark to capture network traffic samples containing both normal and attack flows. The normal flows were synthesized by emulating activities like web browsing, file transfers, and email exchanges. The malicious flows were generated by simulating real-world threats such as malware command channels, port scans, brute force and DDoS.
Below is the description of each attack:}
\textcolor{black}{
\begin{itemize}
\item \textbf{Botnet Attack}: We established a controlled network environment with typical traffic interactions involving five hosts. As part of the simulation, Command and Control (C2) communications from two botnet-infected hosts were introduced. 
\item \textbf{Brute Force Attack}: Emulating the real-world scenario of user logins, a series of user login attempts on an SSH server were simulated. The goal was to observe how effectively a model can discern between legitimate and malicious login attempts. 
\item \textbf{Port Scan}: In a setting that emulates standard traffic to web and database servers from five distinct hosts, port scanning activities were simulated. One host was tasked with initiating port scans to probe network services. Concurrently, two hosts attempted brute force password guesses, enhancing the complexity of the environment. 
\item \textbf{DDoS Attack}: To replicate the havoc wreaked by Distributed Denial of Service (DDoS) attacks, our setup included two web servers catering to normal user traffic. Simultaneously, a barrage of flooding traffic from ten bots was unleashed, aiming to incapacitate the servers. 
\end{itemize}
}
\textcolor{black}{
We utilized CICFlowMeter to extract over 68 statistical flow-level features from the raw traffic, including packet counts, byte counts, jitter, flow durations, and other distributions. By evaluating the IDS on these comprehensive traffic features, we can assess its performance on informative traits seen in operational environments. 
To quantify performance against adversarial evasion attempts, we devised a iterative optimization attack approach that modifies key features to mimic normal behavior while preserving malicious functionality. We tested attack success rates across different perturbation budgets.}

\textcolor{black}{
For each of these attacks, we collected a balanced dataset comprising 5,000 normal traffic flows and 5,000 malicious flows. The detection rate, defined as the proportion of malicious flows correctly identified, served as our primary metric of performance.
It's noteworthy that our simulation constrained the attackers, allowing them knowledge of only 10\% of the dataset and 50\% of the features. Despite this limitation, the results were compelling.
From Table~\ref{tab_real}, we observed the following outcomes post-attack: The initial detection rate of 92\% plummeted drastically to 12\% for Botnet Attack. This significant drop highlights the model's vulnerability to botnet intrusions even when attackers have limited knowledge. More pronounced was the decrease in the detection rate for brute force attacks. Originally boasting a 95\% detection rate, post-attack measurements revealed a sharp decline to a mere 5\%.
Overall, our research on an operational testbed underscores the importance of rigorous adversarial testing to ensure NIDSs are dependable in real-world deployment. We plan to continue expanding the diversity of data, features, and attacks evaluated.
}

\begin{table}
     \caption{Evaluations in Real Environment (Detection Rate)}  
     \label{tab_real}
     \centering
\begin{tabular}{@{}lllll@{}}
\toprule
Stage       & Botnet & Brute Force & Port Scan & DDoS \\ \midrule
original       & 92\%   & 95\%        & 97\%      & 98\% \\
post-attack & 12\%   & 5\%         & 13\%      & 16\% \\ \bottomrule
\end{tabular}
\end{table}

\section{discussion}
\label{sec-discuss}
In this section, we discuss some limitations and potential future directions of ETA.

\subsection{Limitation}
First, our method of crafting AEs is not a direct modification of the attack packets sequence.
We utilize constrained feature-space attacks to mimic traffic-space attacks.
However, for some feature extractors (e.g., AfterImage), it is difficult to transform traffic-space constraints into inverse feature-mapping problems in feature-space.
Second, the features of NIDSs classifiers have various costs to be manipulated. Costs between features are asymmetrical and cannot be accurately captured by our cost models based on the $L_p$-norm.  
Another limitation is that our adversarial attacks can not achieve a high attack success rate for anomaly detection models.

\subsection{Future Work}
We perceive that the module to generate real adversarial traffic is a parallel research topic and hence is not the focus of our study here.
And future work can look into making a formulaic correlation between the feature-space and the traffic-space.
Moreover, a more exciting direction is to learn domain constraints directly from data and then to integrate the learned constraints into the AEs crafting process.
We will seek a new approach to translate domain knowledge about features into cost-driven constraints to address the second limitation.
In terms of the third limitation, we will continue to explore the essence of adversarial transferability, then improve the attack success rate of adversarial attacks against anomaly detection models.
Lastly, while our research is focused on ML-based NIDSs, our framework can be applied to other similar security applications in the future.

\section{Conclusion}
\label{sec-conclusion}
In this work, we are the first to conduct an extensive study of the transfer-based adversarial attack on ML-based NIDSs.
Specifically, we develop a novel transfer-based attack method based on an ensemble substitute model and Important-Sensitive Feature Selection (ISFS).
Moreover, we present a combination of cooperative game theory and perturbation interpretation better to understand the reason for AEs and adversarial transferability.
Evaluation results show that ETA-based systems can craft AEs with high transferability and interpret the behaviors of AEs with high quality.
Additionally, we uncover two significant misconceptions about applying machine learning technology to NIDSs. 
These findings could provide security researchers with an ideal direction for developing ML-based NIDSs, considering model robustness and keeping features at the forefront.

\normalem  
\bibliographystyle{IEEEtran}
\bibliography{references}


\end{document}